\documentclass[fleqn,usenatbib]{mnras}

\usepackage{newtxtext,newtxmath}

\usepackage[T1]{fontenc}

\DeclareRobustCommand{\VAN}[3]{#2}
\let\VANthebibliography\thebibliography
\def\thebibliography{\DeclareRobustCommand{\VAN}[3]{##3}\VANthebibliography}

\usepackage{graphicx}	
\usepackage{amsmath}	

\usepackage{amsmath}
\usepackage{float}
\usepackage[usenames,dvipsnames]{xcolor}

\usepackage{ulem}

\usepackage{graphicx}
\newcommand{\orcid}[1]{
  \href{https://orcid.org/#1}{\includegraphics[scale=1]{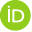}}
}

\renewcommand{\leq}{\leqslant}  \renewcommand{\le}{\leqslant}
\renewcommand{\geq}{\geqslant}  \renewcommand{\ge}{\geqslant}
\newcommand{\pow}[1]{^{#1}}
\newcommand{\poww}[1]{\pow{2}}
\newcommand{\powww}[1]{^{3}}
\newcommand{\sub}[1]{_{#1}}

\newcommand{\half}{{\textstyle\frac{1}{2}}}

\newcommand{\bs}[1]{\boldsymbol{#1}}

\renewcommand{\P}{{\mathbfit{P}}}

\newcommand{\vdot}{{\mathbf{\cdot}}}
\newcommand{\vcross}{{\mathbf{\times}}}
\newcommand{\grad}{\mbox{\boldmath$\nabla$}}

\newcommand{\thth}{\hspace{1.5pt}}
\newcommand{\curl}{\grad\vcross}

\renewcommand\div{\grad\vdot}

\newcommand{\dgrad}[1]{#1\vdot\bs{\nabla}}

\newcommand{\alf}{Alfv{\'e}n}

\newcommand{\bv}{Brunt-V\"ais\"al\"a}

\let\oldeqref\eqref
\renewcommand{\eqref}[1]{Eq. \oldeqref{#1}}

\newcommand{\abs}[1]{\left| #1 \right|}

\newcommand{\lrp}[1]{\left( #1 \right)}
\newcommand{\lrb}[1]{\left[ #1 \right]}

\title[Hall-Induced Magnetoacoustic to Alfv{\'e}n Conversion]{Benchmarking Hall-Induced Magnetoacoustic to Alfv{\'e}n Mode Conversion in the Solar Chromosphere}
\author[Raboonik \& Cally]{
Abbas Raboonik$^{}$\thanks{E-mail: abbas.raboonik@monash.edu}\orcid{0000-0002-6408-1829},
Paul S.~Cally$^{}$\thanks{E-mail: paul.cally@monash.edu}\orcid{0000-0001-5794-8810}
\\
$^{}$School of Mathematics and Monash Centre for Astrophysics, Monash University, Melbourne, Victoria 3800, Australia\\
}

\date{Accepted 2021 August 4. Received YYY; in original form ZZZ}

\pubyear{2021}

\begin{document}
\label{firstpage}
\pagerange{\pageref{firstpage}--\pageref{lastpage}}
\maketitle
\begin{abstract}
A 2.5D numerical model of magnetoacoustic-Alfv\'en linear mode conversions in the partially ionised low solar atmosphere induced by the Hall effect is surveyed, varying magnetic field strength and inclination, and wave frequency and horizontal wave number. It is found that only the magnetic component of wave energy is subject to Hall-mediated conversions to Alfv\'en wave-energy via a process of polarisation rotation. This strongly   boosts direct mode conversion between slow magneto\-acoustic   and Alfv\'en waves in the quiet low chromosphere, even at mHz frequencies. However, fast waves there, which are predominantly acoustic in nature, are only subject to Hall-  induced conversion via an indirect two-step process: (i) a geometry-induced fast-slow transformation near the Alfv\'en-acoustic equipartition height $z_{\rm eq}$; and (ii) Hall-rotation of the fast wave in $z>z_{\rm eq}$. Thus, for the two-stage process to yield upgoing Alfv\'en waves, $z_{\rm eq}$ must lie below or within the Hall-effective window $0\lesssim z\lesssim700$ km. Magnetic field strengths over 100 G are required to achieve this. Since the potency of this Hall effect varies inversely with the field strength but directly with the wave frequency, only frequencies above about 100 mHz are significantly affected by the two-stage process. Increasing magnetic field inclination $\theta$ generally strengthens the Hall   convertibility, but the horizontal wavenumber $k_x$ has little effect.   The direct and indirect Hall mechanisms both have implications for the ability of MHD waves excited at the photosphere to reach the upper chromosphere, and by implication the corona.
\end{abstract}




\section{Introduction}
The solar atmosphere is dominated by magnetohydrodynamic oscillations \citep{BanErdOli07aa}. Many studies have highlighted the potential significance of MHD waves in transport of non-thermal energy, and their contribution to sustaining the temperature structure of the solar atmosphere \citep{Alf47aa, Sch48aa, Ost61aa, De-ErdDe-05aa, Matsumoto2010, Morton2012}. However, the ultimate source and the exact energy-distribution mechanism of these oscillations are yet to be unveiled.

The three classic MHD wave types -- slow, intermediate (or Alfv\'en) and fast -- behave differently as they propagate up from the solar photosphere through the chromosphere and towards the corona. The fast and slow, collectively known as the magneto\-acoustic waves, are both compressive to various extents at various levels. In the high-$\beta$ photosphere and lower chromosphere, the fast wave is essentially acoustic in nature, and readily shocks and thereby heats the atmosphere locally. Depending on where the Alfv\'en-acoustic equipartition level sits, the acoustic (erstwhile fast) wave may reach it in a shocked or unshocked state. In either case though, it splits into distinct acoustically-dominated (now slow) and magnetically dominated (now fast) components in the overlying low-$\beta$ upper chromosphere. In the case of shocks, the resulting slow wave is smoothed and soon re-shocks, but the fast wave remains a shock \citep{PenCal21dv}. However, the fast wave in the upper chromosphere is strongly reflected by the steep Alfv\'en speed gradient, or the transition region if it reaches it, and so contributes little directly to the corona \citep{SriBalCal21vr}. This leaves the Alfv\'en wave to do the `heavy lifting' \citep{MatJesErd13aa}.

Since its discovery by \cite{Alf42aa}, many observational and theoretical studies have been launched to corroborate the existence and perhaps preponderance of \alf{} waves in the upper atmosphere and explain the extreme temperature of the corona \citep{Alf47aa,Kuperus1981,Poedts2002,Cranmer2005,De-ErdDe-05aa,De-McICar07aa, TomMcIKei07aa}. While some works concentrate on uninterrupted transmission of energy by \alf{} waves from photospheric heights all the way to the corona \citep{Cranmer2005}, others have been investigating the significance of mode converted \alf{} waves elicited from originally magnetoacoustic oscillations \citep{CalGoo08aa,Cal17aa,CalKho15aa,CalKho18aa,Gonzalez-Morales2019, Raboonik2019}.  

In ideal MHD, magnetoacoustic oscillations can undergo mode transformation where the local sound speed matches the radially variant \alf{} speed \citep{SchCal06aa}. This may occur even in 2D. However, ideal MHD \alf-magnetoacoustic    transformations can only happen in 3D \citep{CalGoo08aa, CalHan11aa}. On the other hand, non-ideal MHD mode   conversion  mechanisms pervade the partially ionised solar photosphere/lower chromosphere. 
Amongst these   processes, the Hall effect is of particular interest in that it exerts a conservative force \citep[for a full discussion on Hall MHD see][]{Pandey2008}, leading to lossless conversions of energy. Thus, even compressive forms of wave-energy can potentially wind up as \alf{} waves higher up in the atmosphere. 

In a magnetically driven fluid, any kind of fluctuation seeds MHD waves. Thus, it is no surprise that the highly energetic small- and large-scale granulation arising from convective motions in the photosphere have been suspected as viable wave drivers with enough kinetic energy to fuel the hot corona. \cite{WitNoy77aa} estimate the total coronal energy loss to be of order of $10^{5}\,\mathrm{erg}\,\mathrm{cm}^{-2}\,\mathrm{s}^{-1}$ in the quiet sun and coronal holes, and $10^{7}\,\mathrm{erg}\,\mathrm{cm}^{-2}\,\mathrm{s}^{-1}$ in active regions.

Premised on the theory of sound generation developed by \cite{Lighthill1952}, \cite{MusRosSte94aa} show that subsonic photospheric convective motions in the Sun can trigger the production of acoustic waves of about $5 \times 10^{7}\,\mathrm{erg}\,\mathrm{cm}^{-2}\,\mathrm{s}^{-1}$ with a frequency peak around $100\,\mathrm{mHz}$.
Thus, if transportable, this would be more than enough energy required to heat the solar corona. 

In a different study, \cite{Cranmer2005} isolate transverse waves by modelling the solar granular motions as a non-rotational stochastic velocity field and use it as boundary conditions to excite `Alfv\'enic' waves at the photosphere. Their model estimates a total wave-energy flux density of about $10^8 \,\mathrm{erg}\,\mathrm{cm}^{-2}\,\mathrm{s}^{-1}$ at the photosphere (see Figure~12 in that paper). The transverse oscillations are then traced all the way to interplanetary space and dissipated using several nonlinear turbulent damping processes. Their results shows comparable heating rates to empirical data measured at $2R_\odot$.

However, high-$\beta$ transverse oscillations -- which \cite{Cranmer2005} refer to as `\alf{}ic' waves -- are a mixture of \alf{} and magnetic slow waves, depending on their local polarisation. That is, \alf{} waves are polarised along $\bs{k}\vcross\bs{B}$ where $\bs{k}$ is the wavevector as defined in the WKB approximation at sufficiently high frequencies, while high-$\beta$ slow waves oscillate in the direction of $\bs{k}\vcross\lrp{\bs{k}\vcross\bs{B}}$.
Part of the present study is motivated by the prospect of obtaining upgoing \alf{} waves out of input slow waves at the base.

The remainder of the paper is laid out as follows. first, an exposition of the model including the choice of the boundary conditions necessary to ensure exclusive excitation of slow waves is presented in Section~\ref{sec:model}. We then proceed to presenting the results in Section~\ref{sec:results}. Finally, we conclude the paper with a discussion in Section~\ref{sec:summary}.

\section{Mathematical Model}\label{sec:model}
Our setup consists of the plane-stratified quiet Sun atmospheric Model C of \cite{VerAvrLoe81aa}, permeated by a uniform 2D magnetic field of selectable inclination $\theta$ from the vertical, prescribed by $\bs{B}_0 = B_0 (\sin\theta, 0, \cos\theta)$. The model covers a solar region including the partially ionised photosphere up to mid chromosphere, spanning from $z_{\rm bot} = -0.5$ Mm to $z_{\rm top} = 1.6$ Mm, to which we will refer as `the main box'. The interval over which the Hall effect is of major significance sits inside the partially ionised patch, and in line with \cite{CalKho15aa}, will be called the `Hall window'.

The motivation here is to set up  numerical MHD-wave dispersion experiments as they pass through a gravitationally stratified semi-realistic solar setting to   isolate the role of the Hall effect, and determine   its efficacy in mode converting high-$\beta$ slow/fast waves into upgoing \alf{} waves. The phrase `semi-realistic' implies that despite the employed realistic model atmosphere, appropriate artificial isothermal boundaries are imposed to craft pure slow or fast waves at the base, and annihilate any incoming waves at the top. Even though the enforced isothermality condition may be somewhat justifiable at $\beta \gg 1$, it most certainly does not capture the essence of low-$\beta$ regions.  Therefore, we entirely ignore the impacts of the transition region and the corona. The dynamic and highly structured nature of the chromosphere is also ignored.

Upon entering the Hall window, an input slow or fast wave encounters two possibilities depending on the location where the sound and Alfv\'en speeds $c_s$ and $a$ are equal ($\beta=2/\gamma$). This special location is known as the equipartition level $z_{\rm eq}$. If $z_{\rm eq}$ is situated in the Hall window, then the wave can undergo another conversion induced by geometric effects, which may have vital implications in the production of \alf{} waves. This process is referred to as a two-stage conversion and is addressed in Section~\ref{subsec:twostage}.

To set the scene, we artificially augment the main box along the $z$-axis by superimposing two isothermal slabs (of the same local temperature) at both ends. This enables us to use the exact 2.5D isothermal solutions of \cite{HanCal09aa} within the isothermal patches and generate pure slow or fast oscillations as desired.

\begin{figure}
    \centering
    \includegraphics[width=0.4\textwidth]{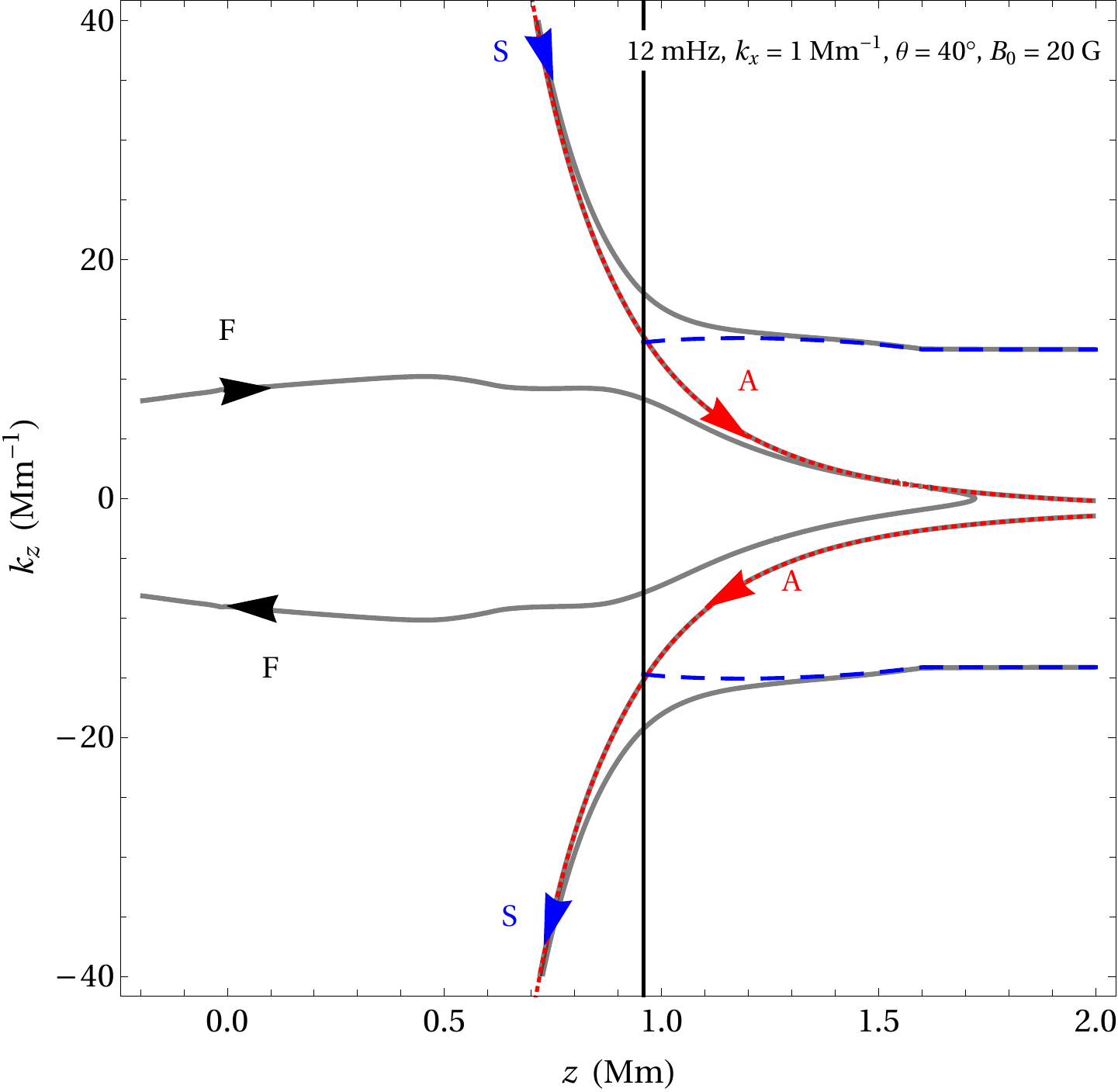}
    \caption{Dispersion diagram of the system at 12 mHz. Each letter next to the arrows represents the first letter of the three classes of MHD waves, namely, `Fast', `Slow', and `\alf{}', with the \alf{} mode highlighted in red. The solid vertical line indicates the equipartition level `$z_{\rm{eq}}$'. The dashed blue curves represent the magnetic part of the magnetoacoustic modes. As can be seen, the curves corresponding to the slow and the \alf{} waves run very closely over an extended region, allowing for potential perpetual mode transformations between these modes. Past `$z_{\rm{eq}}$', the slow mode becomes magnetic, the fast wave reflects back,
    and the \alf{} wave transmits through uninterruptedly.}
    \label{fig:dispersion}
\end{figure}

To get a better perspective on mode conversions in MHD, dispersion diagrams offer revealing insights into the behaviour of individual modes in $z$--$k_z$ phase space. 
Figure~\ref{fig:dispersion} depicts the ray trajectories of the three MHD modes at 12 mHz, where   clearly distinct wave types (based on their loci in the diagram) are labelled by the first letter of their names   \citep[-- Alfv\'en, Fast and Slow -- see equation~(22) of][and the dispersion diagrams therein]{Raboonik2019}.
The solid line signifies the equipartition level. As can be seen, the curves belonging to the slow and the \alf{} waves are nearly coincident up to $z_{\rm eq}$, whereas the upgoing fast and \alf{} trajectories come very close only past $z_{\rm eq}$.
These extended regions of adjacency indicate similar asymptotic behaviours between the participant waves and provide the capacity for mode conversions to take place. However, whether or not any transformation will actually transpire comes down to availability of    conversion  mechanisms in the system. 

In the absence of partial ionisation processes such as the Hall effect, the magnetoacoustic-\alf{}    transformation can only occur in full 3D, provided the magnetic field $\bs{B}_0$, gravity $\bs{g}$ and the wave vector $\bs{k}$ are not coplanar. However, upon aligning these vectors into the same plane while allowing for the oscillations to develop in 3D, the system takes on the so-called 2.5 dimensionality. Thence, owing to its unique polarisation, the \alf{} wave polarised in the perpendicular direction would decouple identically from the magnetoacoustic waves that are polarised entirely in the plane. Thus, despite the mode transformation potentiality of the system as suggested by Figure~\ref{fig:dispersion}, there would be no way for the magnetoacoustic oscillations to communicate and convert to \alf{} waves and vice versa. Nevertheless, the slow and the fast waves can still mode convert to one another at $z_\mathrm{eq}$ \citep{SchCal06aa}.

On the other hand, once partial ionisation is activated, the Hall effect can furnish a unique inter-modal communication means for magnetoacoustic-\alf{}    conversions  \citep{CalKho15aa}. In this paper, $\bs{B}_0$ and $\bs{k}$ are confined in the $x-z$ plane, hence any oscillation along the $y$-axis is readily identified as the \alf{} type, whereas the polarisation vectors of the pure fast and slow waves will be confined in the same plane as $\bs{B}_0$.

There are two significant nuances in the present study in contrast with \cite{Raboonik2019}. First, the incorporated realistic anisothermal atmospheric model, and second, here we dispense with the `small Hall parameter $\epsilon_H$' assumption which was crucial to the semi-analytical solutions in that paper. The latter improvement allows the waves to remain    transformable across the conversion zone and continuously exchange energy in a feedback loop.

It is easy to think that the distinction between slow and Alfv\'en waves is moot, given their very similar asymptotic behaviours at high plasma-$\beta$. However, Fig.~\ref{fig:dispersion} suggests that they are very different globally. In the absence of any mode conversion, the slow and Alfv\'en loci are very different around and above $z_{\rm eq}$, with implications for upper atmospheric heating. Any    conversion  at low altitudes between slow and Alfv\'en therefore has global implications.

\subsection{Governing Equations}\label{sec:gov}
\begin{figure}
    \centering
    \includegraphics[width=0.48 \textwidth]{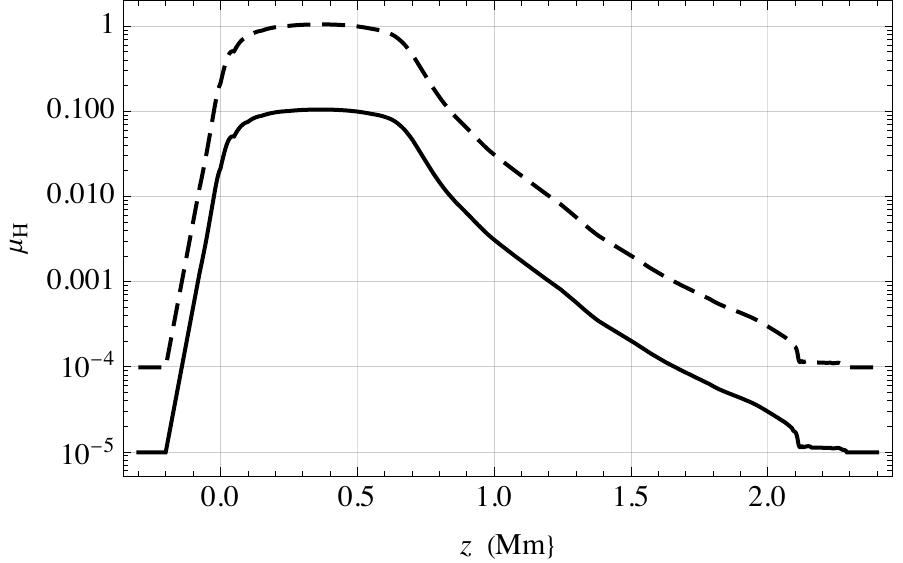}
    \caption{The isothermally extended profile of the Hall parameter $\mu_H$ in SI units versus height in logarithmic scale for $B_0=10$ G (full curve) and 100 G (dashed). The magnitudes of $\mu_H$ in the isothermal tails where the numerical boundaries sit are negligible, and hence the Hall term is effectively switched off in those patches.}
    \label{fig:muH}
\end{figure}

Assuming time-homogeneity in the time interval of interest, i.e., $\partial/\partial t = -i\omega$, and a static equilibrium state, i.e., $\bs{v}_0 = 0$, the set of linearised Lagrangian MHD equations governing the system in the Cartesian coordinates is given by,
\begin{subequations}\label{eq:mhd}
\begin{equation}\label{eq:momentum}
     -i\,\omega\,\rho_0\bxi = -\grad{p_1} + \rho_1 \bs{g} + \bs{j}_1\vcross\bs{B}_0
\end{equation}
\begin{equation}\label{eq:mass}
    \rho_1 = -\div{\lrp{\rho_0 \bxi}}
\end{equation}
\begin{equation}\label{eq:induction}
    \bs{b} = \curl\left(\bxi\vcross\bs{B}_0 - \frac{i \mu_0 \mu_H}{\omega\rho_0 B_0} \bs{j}_1\vcross\bs{B}_0\right)
\end{equation}
\begin{equation}\label{eq:amper}
    \bs{j}_1 = \frac{1}{\mu_0}\curl{\bs{b}}
\end{equation}
\begin{equation}\label{eq:energy}
    p_1 = -\dgrad{\bxi} p_0 - \gamma p_0 \div{\bxi},
\end{equation}
\end{subequations}
where $\bxi$ is the vector plasma displacement, $\bs{b}$ is the perturbed magnetic field, $\mu_0$ is the magnetic permeability in vacuum, $\mu_H = \rho_0 B_0/\mu_0 n_e e$ is the Hall parameter (plotted in Fig.~\ref{fig:muH} for two magnetic field strengths), and the symbols indexed with zero and one indicate the equilibrium and the perturbed quantities, respectively. Now, on account of the 2.5D setup and the symmetry of the system across the $x-y$ plane, we have
\begin{equation}\label{eq:displacement}
    \bxi = \lrp{\xi(z), \eta(z), \zeta(z)} e^{i (k_x x - \omega t)}.
\end{equation}
 Combining equations~(\ref{eq:mhd}) and (\ref{eq:displacement}) into a single vectorial differential equation, one arrives at a set of three second order ODEs that fully describe the behaviour of $\xi$, $\eta$, and $\zeta$ all over the main box (see Appendix~\ref{sec:app2} for a derivation).

The significance of the Hall effect at any particular position and wave frequency is best assessed via the dimensionless `Hall parameter' $\epsilon_H=\omega/f_{\rm i}\Omega_{\rm i}=\omega\,\mu_H/\rho a^2$, where $\Omega_{\rm i}=Z e B_0/m_{\rm i}$ is the ion cyclotron frequency, $f_{\rm i}$ is the ionisation fraction, $m_{\rm i}$ is the mean ion mass, $Z$ is the ion mean charge state, and $e$ is the elementary charge. Note in particular that it increases linearly with wave frequency, and varies inversely with magnetic field strength (since $\mu_H\propto B_0$ and $a^2\propto B_0^2$).

\subsection{Boundary Conditions}\label{sec:BCs}
To close the system, we need to impose suitable boundary conditions to inject the system with a pure slow, fast or Alfv\'en wave from below and disallow any   other incoming waves   at top or bottom. As briefly touched on before, we artificially extend the main box by replicating the end-values of all the atmospheric variables down to $z = -0.887$ Mm at the bottom and up to $z = 7.11$ Mm at the top\footnote{This is to ensure that the boundary slabs are far enough from the regions of fast/slow, fast/Alfv\'en and slow/Alfv\'en mode    conversion  that it is possible to identify the separate modes by their asymptotic behaviours in the limits $z\to\pm\infty$}., and place the numerical boundaries at these new  locations.
  The extension slabs are chosen to be isothermal so that we may use the exact solutions of \cite{HanCal09aa} to set up the boundary conditions.

  The exact solutions in the isothermal slabs may be expressed in terms of $_2F_3$ generalised hypergeometric functions \citep{Luk75aa}, using the magnetic field coordinate system $(\hat{\bs{e}}_{\perp}, \hat{\bs{e}}_{\parallel}, \hat{\bs{e}}_y)$, where $\perp$ and $\parallel$ indicate perpendicular and parallel to $\hat{\bs{B}}_0$ in the $x-z$ plane, respectively, and $\hat{\bs{e}}_y$ remains the same as the unit polarisation vector of the \alf{} wave oriented along the $y$-axis.
  These are entirely equivalent to the Meijer-G function solutions of \cite{ZhuDzh84aa}, but more convenient to use. The exact isothermal mixed-mode solution (to equation~(\ref{eq:xiperp})) for magnetoacoustic oscillations is given by 
\begin{align}\label{eq:chiperpFull}
\xi_{\perp}(s) =\, &  C_{1}\, s^{-2 \kappa } \, _2F_3[-\kappa -i \kappa_{z}+\frac{1}{2},-\kappa +i \kappa_{z}+\frac{1}{2};1-2 \kappa ,    \nonumber \\
 &\quad -i\kappa \tan \theta-\kappa -i \kappa_{0}+\frac{1}{2},-i \kappa \tan \theta  -\kappa +i \kappa_{0}+\frac{1}{2}; \nonumber \\
 &\quad-s^2 \sec ^2 \theta]    \nonumber\\
 + &~C_{2}\, s^{2 \kappa } \, _2F_3[\kappa -i
   \kappa_{z}+\frac{1}{2},\kappa +i \kappa_{z}+\frac{1}{2};2 \kappa +1,      \nonumber\\
 &\quad -i\kappa \tan \theta +\kappa -i
   \kappa_{0}+\frac{1}{2},-i\kappa \tan \theta  +\kappa +i \kappa_{0}+\frac{1}{2};\nonumber \\
  &\quad-s^2 \sec ^2 \theta]    \nonumber\\
  + &~C_{3}\, s^{1+i (2 \kappa  \tan \theta-2 \kappa_{0})} \, _2F_3[-i \kappa_{0}-i
   \kappa_{z}+i \kappa  \tan \theta+1,    \nonumber\\
 &\quad -i \kappa_{0}+i \kappa_{z}+i \kappa  \tan \theta+1;1-2 i
   \kappa_{0}, \nonumber\\
  &\quad i\kappa \tan \theta  -\kappa -i \kappa_{0} +\frac{3}{2}, i\kappa \tan \theta  +\kappa -i
   \kappa_{0}+\frac{3}{2};-s^2 \sec ^2 \theta]    \nonumber\\
   + &~C_{4}\, s^{1+i (2 \kappa  \tan \theta+2 \kappa_{0})} \, _2F_3[i \kappa_{0}-i \kappa_{z}+i \kappa  \tan \theta +1,  \nonumber\\
  &\quad i \kappa_{0}+i
   \kappa_{z}+i \kappa  \tan \theta + 1;2 i \kappa_{0}+1, i\kappa \tan \theta-\kappa +i \kappa_{0}+\frac{3}{2},     \nonumber\\
 &\quad i\kappa \tan \theta  +\kappa +i \kappa_{0}+\frac{3}{2};-s^2 \sec ^2 \theta]   \nonumber\\
 = &  \sum\sub{i = 1}\pow{4} C\sub{i}\,\xi_{\perp,i}(s).
\end{align}
Here, $C_i$ are arbitrary constant coefficients, and we have defined the following dimensionless parameters
\begin{gather}
    \nu = \frac{\omega H}{c_s} \label{eq:nu}\\
    s = \nu e^{-z/2H} \\
    n = \frac{\sqrt{\gamma - 1}}{\gamma} \\
    \kappa = k_x H \\
    \kappa_0 = \sqrt{\nu^2 \sec^{2}\thth\theta - \frac{1}{4}} \\
    \kappa_z = \sqrt{\nu^2 + \frac{(n^2 - \nu^2)\kappa^2}{\nu^2} - \frac{1}{4}} \label{eq:kapaz},
\end{gather}
in which $H$ is the density scale height, $s$ is the dimensionless measure of height, $\gamma$ is the heat capacity ratio, and $n$ is the \bv{} frequency. 
  The three $\kappa$ coefficients are dimensionless wavenumbers that naturally arise in the analysis of non-magnetic acoustic gravity waves.
Equation~(\ref{eq:chiperpFull}) together with equation~(\ref{eq:xipar}) fully describe the behaviour of $\xi$ and $\zeta$ in the isothermal boundary slabs, and will be analyzed in section~\ref{subsub:magacBoundaries} to find suitable choices for the $C_i$. The magnetoacoustic displacement components in the Cartesian and magnetic field coordinates are related through $\xi_\parallel = \xi \sin\thth\theta + \zeta \cos\thth\theta$ and $\xi_\perp = \xi \cos\thth\theta - \zeta \sin\thth\theta$.

  On the other hand, the behaviour of the \alf{} wave in the 2.5D isothermal regime outside the Hall-window (governed by equation~(\ref{eq:eta})) is conveniently given by types 1 and 2 of the Hankel functions of zeroth order, according to
\begin{equation}\label{eq:alf}
    \eta(s) = s^{2 i \kappa \tan \theta} \left(D_{1} H_0^{(1)}(2 s \sec \theta) + D_{2} H_0^{(2)}(2 s \sec \theta) \right),
\end{equation}
where $D_i$ are arbitrary constants corresponding respectively to downgoing and upgoing Alfv\'en waves.

 
\subsubsection{Boundary conditions on $\eta$}
As shown in Figure~\ref{fig:muH}, the Hall parameter is negligible anywhere outside the interval $-0.23$ Mm$\, < z < 2.2$ Mm. Thus, within the boundary slabs, $\eta$ obeys equation~(\ref{eq:alf}).
Therefore, to eliminate incoming \alf{} waves at the boundaries, we must choose $D_1 = 0$ at the top, and $D_2 = 0$ at the bottom.

  \subsubsection{Boundary conditions on $\xi$ and $\zeta$}\label{subsub:magacBoundaries}
Away from $z_{\rm eq}$, the mixed-mode solution for magnetoacoustic oscillations given by equation~(\ref{eq:chiperpFull}) can be asymptotically broken down into slow and fast wave contributions.

At the top boundary where $s\to0$ (i.e., $z \to\infty$),   equation~(\ref{eq:chiperpFull}) takes on the asymptotic form
\begin{multline}\label{chiperp}
    \xi_{\perp,\rm Top}(s) \sim  C_{1} s^{-2 \kappa } + C_{2} s^{2 \kappa } + C_{3}\, s^{1+i (2 \kappa  \tan \theta-2 \kappa_{0})} \\
 +\, C_{4}\, s^{1+i (2 \kappa  \tan \theta+2 \kappa_{0})},
\end{multline}
from which the solution to $\xi_{\parallel}$ follows readily by solving equation~(\ref{eq:xipar}). Assuming $\kappa > 0$, the individual terms in equation~(\ref{chiperp}) in order from left to right correspond to, the exponentially growing fast mode (nonphysical), the exponentially decaying fast mode, the upgoing slow mode, and the downgoing slow mode. As stated earlier, for the purposes of this study we immediately set $C_1 = C_4 = 0$, to dispense with the nonphysical solution and the incoming slow wave at the top.

At the bottom boundary where $s \to\infty$ (i.e., $z \to-\infty$), each of the four solutions couple to four leading order asymptotic behaviours according to,
\begin{multline}\label{eq:chiperpbot}
    \xi_{\perp, j}  \sim a_{1j} s^{-1/2+2i\kappa\tan\theta} \text{e}^{2is\sec\theta}
    + a_{2j} s^{-1/2+2i\kappa\tan\theta} \text{e}^{-2is\sec\theta} \\
    + a_{3j} s^{-1+2i\kappa_z} + a_{4j} s^{-1-2i\kappa_z},
\end{multline}
whose first and second terms are identified as the downgoing and the upgoing slow waves, respectively. However, the remaining two terms may be interpreted depending on the position of the system in the acoustic-gravity diagram as presented in Figure~\ref{fig:regions}. In Region I, where waves are essentially driven by the gas pressure, these terms represent the downgoing and the upgoing fast waves, respectively, whereas the propagation directions are reversed in Region II, where gravity dominates. Ultimately, all waves are evanescent in regions III and IV and hence carry no energy. Therefore, the general solution at the bottom boundary can be written in terms of these behaviours as
\begin{multline}\label{eq:asymBot}
    \xi_{\perp,\rm Bot}(s) \sim c_1\, s^{-1/2+2i\kappa\tan\theta} \text{e}^{2is\sec\theta}
    + c_2\, s^{-1/2+2i\kappa\tan\theta} \text{e}^{-2is\sec\theta} \\
    + c_3\, s^{-1+2i\kappa_z} + c_4\, s^{-1-2i\kappa_z},
\end{multline}
where the vector coefficients $\bs{C} = (C_1, C_2, C_3, C_4)^T$ and $\bs{c} = (c_1, c_2, c_3, c_4)^T$ connect by
\begin{equation}\label{eq:cac}
    \bs{c} = \bs{A}\, \bs{C},
\end{equation}
in which $\bs{A}=(a_{ij})$ is a known $4\times4$ matrix \citep{HanCalDon16aa}. 

Thus, setting $c_4 = 0$ in Region I and $c_3 = 0$ in Region II, to block out the upgoing fast wave at the bottom boundary, and $c_1 = 1$ to inject it with a normalised slow wave, equation~(\ref{eq:cac}) can be used to determine the remaining coefficients $c_i$.
These coefficients determine how much of the incident slow wave will reflect, mode convert to fast, and transmit through the barrier of the equipartition level. Similar considerations can be employed to craft pure normalised fast waves instead.

 Having set up the boundary conditions, it remains to solve equations~(\ref{eq:mhd}) within the main box and find the energy fluxes corresponding to each   asymptotically decoupled (or locally weakly coupled) slow, fast, and \alf{} oscillations that passes through the end-planes. For this task, the equations are first combined and simplified to a set of three coupled ODEs in $\bxi$ as listed in equations~(\ref{eq:mainPDEs}) (see Appendix~\ref{sec:app2}), and then solved numerically using a shooting method.  
 
\subsubsection{Nature of the asymptotic solutions}\label{sec:asymptoticNature}
It is important to distinguish the essence of MHD waves around $z_{\rm eq}$ and far away from it. For instance, at and in the vicinity of $z_{\rm eq}$, 2.5D magnetoacoustic waves have an indivisible mixed-mode characteristic which can only be described by the exact solution given in equation~(\ref{eq:chiperpFull}).
However, as we recede far from $z_{\rm eq}$ in either direction, they begin to locally exhibit simpler behaviours corresponding to the asymptotic formulae (\ref{chiperp}) and (\ref{eq:asymBot}) identifiable as decoupled slow and fast waves. In other words, although the exact ${}_2F_3$ solutions represent global mixed-modes containing both fast and slow characteristics, these behaviours may be decoupled locally in $z\ll z_{\rm eq}$ and $z\gg z_{\rm eq}$. Figure \ref{fig:dispersion} indicates that such a decoupling is well justified in the upper and lower reaches of our computational domain.

It is also of note that the leading order asymptotic behaviour for $\eta$ of the Alfv\'en solutions expressed in equation (\ref{eq:alf}) as $s\to\infty$ are identical to the downgoing and upgoing slow wave asymptotic terms for $\xi_\perp$ of equation (\ref{eq:asymBot}). This near-degeneracy of the Alfv\'en wave $\eta$ and the slow wave $\xi_\perp$ where $a\ll c$ only applies to these components though; the slow wave also exhibits a plasma velocity component parallel to the field, which the Alfv\'en wave does not. Nevertheless, we might expect a similar rate of generation of slow and Alfv\'en waves at the photosphere.
Once again, we stress the importance of dispersion diagrams in comprehending this picture. As shown in Figure~\ref{fig:dispersion}, this near degeneracy manifests itself as the near-matching loci of the slow and \alf{} modes below $z_{\rm eq}$. 

This notion of asymptotic analysis of MHD waves is also applicable more broadly. For example, in fully-ionized 3D MHD where the \alf{} mode is no longer decoupled from magnetoacoustic oscillations due to geometric effects, there are no individual \alf{}, slow, or fast modes in the global sense. Nor there are any known functions with which to formulate the exact global solutions of such a system in any generality. Nevertheless, looking at the dispersion curves \citep[e.g., see Figure~1 of][]{CalHan11aa}, we may still learn about the local asymptotic behaviour of the resultant MHD waves following a dispersion experiment far from $z_{\rm eq}$, where the coupling effects are weak and the waves begin to possess simple-wave characteristics. In particular, these diagrams are very useful in discerning which types and under what circumstances waves reach the upper atmosphere and potentially contribute to upper chromospheric and coronal heating.

\subsection{Energy Flux}
The total vector wave-energy flux is comprised of the pressure work and the Poynting flux vector as follows \citep{Cal01ab,HanCalDon16aa},
\begin{equation}
    \bs{\mathcal{F}} = \mathrm{Re}\lrb{i \omega p_1 \bxi^{*} - \frac{i \omega}{\mu_0}\lrp{\bxi^{*} \vcross \bs{B}_0} \vcross \bs{b}}.
\end{equation}
However, we are only interested in the energy transport in the $z$-direction. Additionally, it is instructive to break down the total energy flux into the acoustic, magnetic, and \alf{} components (energy flux functions hereafter) denoted respectively by, $\mathcal{F}_{\rm ac}$ and $\mathcal{F}_{\rm mag}$, and $\mathcal{F}_{\rm A}$, according to
\begin{subequations}\label{eq:magacalf}
    \begin{equation}\label{eq:ac}
        \mathcal{F}_{\rm ac} = -\omega \, \mathrm{Im}\lrb{p_1 \bxi^*\vdot\,\hat{\bs{z}}}
    \end{equation}
    \begin{equation}\label{eq:mag}
        \mathcal{F}_{\rm mag} = \omega \, \mathrm{Im}\lrb{\frac{1}{\mu_0}\lrp{\xi_{\perp}^{*}\hat{\bs{e}}_{\perp} \vcross \bs{B}_0}\vcross\bs{b}}\vdot\,\hat{\bs{z}}
    \end{equation}
    \begin{equation}\label{eq:a}
        \mathcal{F}_{\rm A} = \frac{\omega}{\mu_0}\mathrm{Im}\lrb{\lrp{\eta^{*}\hat{\bs{e}}_{y} \vcross \bs{B}_0}\vcross\bs{b}}\vdot\,\hat{\bs{z}}.
\end{equation}
\end{subequations}
However, caution must be taken as these are interconnected through the displacement vector $\bxi$, and hence their identity does not remain the same throughout the domain. Nevertheless, they may locally represent the energy flux associated with certain   weakly coupled MHD   waves under specific circumstances in the absence of coupling mechanisms   or far from $z_{\rm eq}$ as explained in section~\ref{sec:asymptoticNature}. For instance, it is well established that in the $\beta \gg 1$ regime, the slow magnetoacoustic wave is almost perfectly of pure magnetic attributes, while the fast wave is characterised as acoustic there. This is reflected in Figure~\ref{fig:fluxes}, and means that the slow and the fast waves tend to decouple to a high degree in sub-photospheric depths. However, as we recede from high-$\beta$ regions, various coupling mechanisms weigh in, rendering these modes as part acoustic and part magnetic.

Now, given the coefficients $\bs{c}$ and $\bs{C}$, which are most conveniently applied for the $\beta > 1$ and $\beta < 1$ sub-domains respectively, it is possible to directly calculate the energy flux of the individual MHD modes by
\begin{subequations}\label{eq:fluxes}
    \begin{equation}\label{eq:fluxesa}
        \mathcal{F}_{\rm FS} = \frac{p_{\rm mag} \omega^2}{c_s}
        \begin{cases}
            \bs{C}^{\dagger} \Phi\,\bs{C}, \quad z > z_\mathrm{eq} \\
            \\
            \bs{c}^{\dagger} \Psi\,\bs{c}, \quad z < z_\mathrm{eq} \\
        \end{cases},
    \end{equation}
    \begin{equation}
        \mathcal{F}_{\rm A} = \frac{p_{\mathrm{mag}} \omega^2}{c_s} \frac{\cos^2 \theta}{\pi\nu}\bs{D}^{\dagger}\bs{D}.
    \end{equation}
\end{subequations}
Here $\mathcal{F}_{\rm FS}$ is the magnetoacoustic contribution to the energy flux containing the fast and the slow components, $p_{\mathrm{mag}}$ is the equilibrium magnetic pressure, $\bs{D} = (D_1, D_2)$ (see equation~(\ref{eq:alf})), $\Psi = \bs{A}^{\dagger}\Phi\,\bs{A}$, and following \cite{HanCalDon16aa}
\begin{equation}\label{eq:phi}
    \Phi =\left(
\begin{array}{cccc}
 0 & \frac{i \kappa }{\nu } & 0 & 0 \\
 -\frac{i \kappa }{\nu } & 0 & 0 & 0 \\
 0 & 0 & \phi _{33} \mathcal{U}\left(\kappa _0^2\right) & \phi _{34} \mathcal{U}\left(-\kappa _0^2\right) \\
 0 & 0 & \phi _{34}^* \mathcal{U}\left(-\kappa _0^2\right) & \phi _{44} \mathcal{U}\left(\kappa _0^2\right) \\
\end{array}
\right),
\end{equation}
where $\mathcal{U}$ is the unit step function, and the $\phi_{ij}$ elements can be found in the Appendix of \cite{HanCalDon16aa}.
Equation~(\ref{eq:fluxesa}) implicates that one may isolate the energy flux components due to the fast and the slow waves whenever the association between the $\bs{C}$ elements and the individual magneto\-acoustic modes is known.

Lastly, the outgoing energy fluxes at the boundaries due to each individual   asymptotically decoupled wave can be determined by evaluating equations~(\ref{eq:fluxes}) at $z_{\rm top}$ and $z_{\rm bot}$.
As dictated by the BCs, the energy fluxes at the top can only be due to the outgoing `transmitted' (or `converted') slow and `Hall-induced mode converted' \alf{}  waves. These are denoted by $\mathcal{F}_{\rm S}^{\uparrow}$ and $\mathcal{F}_{\rm A}^{\uparrow}$, respectively.
However, at the bottom, the upgoing slow (or fast) flux is normalised to unity by construction, and hence $\mathcal{F}_{\rm S\uparrow} = 1$ (or $\mathcal{F}_{\rm F\uparrow} = 1$). The remaining fluxes are due to the downgoing `reflected' slow $\mathcal{F}_{\rm S\downarrow}$ (or `converted' fast $\mathcal{F}_{\rm F\downarrow}$), the downgoing `mode converted'  fast $\mathcal{F}_{\rm F\downarrow}$ (or `reflected' slow $\mathcal{F}_{\rm S\downarrow}$), and the Hall-induced mode converted downgoing \alf{} $\mathcal{F}_{\rm A\downarrow}$. Note the difference between the superscript and the subscript arrows. In any case, in view of the normalised injected wave and conservation of energy, we require that
\begin{equation}\label{eq:enecon}
    \lrp{\mathcal{F}_{\rm S}^{\uparrow} + \mathcal{F}_{\rm A}^{\uparrow}} + \lrp{\mathcal{F}_{\rm S\downarrow} + \mathcal{F}_{\rm F\downarrow} + \mathcal{F}_{\rm A\downarrow}} = 1.
\end{equation}

\begin{figure}
    \centering
    \includegraphics[width=0.4\textwidth]{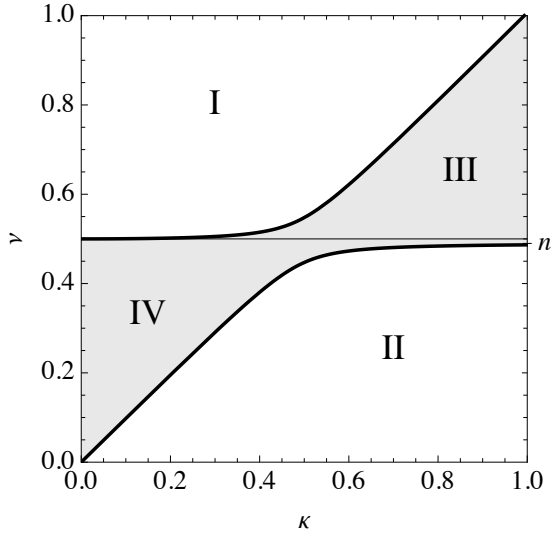}
    \caption{The propagation diagram encapsulating the behaviour of acoustic-gravity waves as they propagate through a stratified non-magnetic isothermal atmosphere. In Region I waves propagate vertically and are predominantly acoustic in nature. In Region II they propagate as essentially gravity waves. In Regions III and IV all waves are evanescent (signified by the grey shading), and so carry no energy vertically. Here $n=\sqrt{\gamma-1}/\gamma=0.4899$ for $\gamma=5/3$ is the dimensionless {\bv} frequency, and $\nu=\half$ is the dimensionless acoustic cutoff frequency.}
    \label{fig:regions}
\end{figure}

\section{Results}\label{sec:results}
We wish to establish a connection between the free parameters $f$, $B_0$, $k_x$, and $\theta$ and the `effectiveness' of the Hall term in generating upgoing \alf{} waves from the input slow waves. Here $f = \omega/2\pi$ is the wave frequency measured in Hertz.
Since the total injected energy flux is normalised to unity, we define this effectiveness simply as the magnitude of the outgoing \alf{} energy flux evaluated at the top, i.e., $\mathcal{F}_{\rm A}^{\uparrow}$. 

The results are presented using (i) contour plots of the local outgoing slow and \alf{} energy fluxes measured at the boundaries of the numerical box, and (ii) plots of the energy flux functions described by equation~(\ref{eq:magacalf}) versus $z$.

\subsection{Pure Slow to \alf{} Conversion}\label{subsec:slow2Alf}
As mentioned above, without the Hall effect we would expect a zero \alf{} energy flux because the injected magneto\-acoustic waves are in the 2D vertical $x$--$z$ plane and there is no geometric coupling mechanism. Nevertheless, magnetoacoustic waves can still communicate and exchange energy between each other in the vicinity of the equipartition level, where they undergo mode transformation, reflection, and transmission through pure resonant coupling.

When the Hall term is switched on, the energy fuelled by the slow wave may partially or entirely convert to an upgoing \alf{} wave depending on the    convertibility strength across the weakly ionised region. Partial transformations occur when the    conversion  effect is either too weak or too strong. In the latter case, conversion overshoots happen where an input slow wave transforms all the way to \alf{} wave and back to slow wave in a cyclic fashion.
In other words, upon entering the Hall window, the slow wave will transmute into a superposition (or inter-modal)   state of part slow- and part \alf-like waves with a rotating polarisation vector replacing its original ($x$--$z$)-plane counterpart.
It is the variable circular speed of this vector that determines the local strength of the Hall    conversion . Hence stronger Hall    conversions lead to more rapidly rotating polarisation vectors, which will manifest themselves as cyclic patterns in the flux diagrams. These cyclic patterns would then eventually cease at the end of the Hall domain, where the    convertibility breaks and the hybrid   wave splits into its `local' amounts of slow and \alf{} wave-energies. 
Therefore, assuming a set of parameters $(\Bar{f}, \Bar{B}_0, \Bar{k}_x, \Bar{\theta})$ that greatly amplifies the strength of the Hall    conversion , even slight shifts in any one of these values can drastically impact the effectiveness. As we shall see, this introduces interesting behaviours in the effectiveness function $\mathcal{F}_{\rm{A}}^{\uparrow}$.

\begin{figure}
    \centering
    \includegraphics[width=0.48\textwidth]{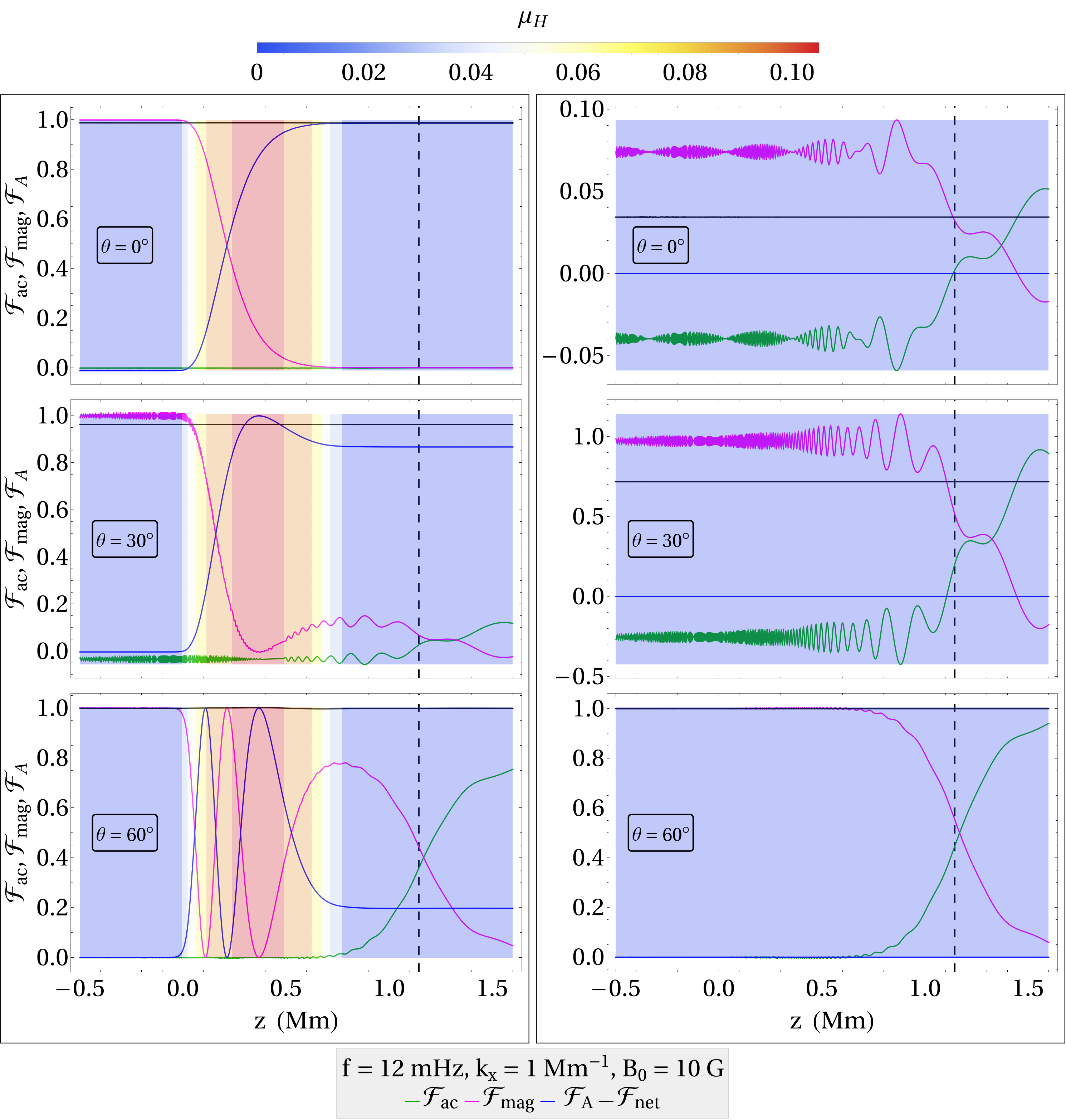}
    \caption{The energy flux functions of an injected slow wave versus height at three varied field inclinations as labelled at 12 mHz. The characteristics of the supplied wave are listed in the legend. The solid horizontal line shows the net upward energy flux and the dashed vertical line flags the position of the equipartition level. The colour-shaded background shows the profile of $\mu_H$. Notice that  all of the injected slow wave-energy in subsurface regions near the bottom boundary is stored in $\bs{\mathcal{F}}_{\rm mag}$, expectedly indicating that the slow wave is predominantly of magnetic nature there.
    \textit{Left column:} The Hall effect is active. At $\theta = 0$, the entire sourced slow wave is mode converted into upward \alf{} wave. The strength of the Hall    conversion  grows with $\theta$, while (in this particular case) the effectiveness declines. The enhancement of the    conversion  effect due to increasing $\theta$ is in part owing to the longer segments of the magnetic field lines that are submerged in the Hall window of the atmosphere. This enables the field-guided waves such as the \alf{} and the high-$\beta$ slow waves to remain    convertible for longer periods and transform into once another.
    \textit{Right column:}  The Hall effect is deactivated, and hence the zero $\bs{\mathcal{F}}_{\rm A}$ curve. At $\theta = 0$, the slow wave is heavily reflected, as indicated by the small positive tail of $\bs{\mathcal{F}}_{\rm mag}$ near the bottom boundary. Past the equipartition level, the fast wave is evanescent and hence carries no energy. Thus, the leftover energy in the $\beta < 1$ region belongs entirely to the transmitted slow wave.}
    \label{fig:fluxes}
\end{figure}

Figure~\ref{fig:fluxes} illustrates the behaviour of the energy flux functions $\bs{\mathcal{F}}_{\rm ac}$ (green), $\bs{\mathcal{F}}_{\rm mag}$ (magenta), and $\bs{\mathcal{F}}_{\rm A}$ (blue) of an input pure slow wave at $f = 12\,\mathrm{mHz}$ versus height for three different magnetic field inclinations. The colour-shaded background represents the profile of the Hall parameter $\mu_H(z)$, with the weakly ionised Hall-dominated patch of the atmosphere sandwiched between the monotone blue portions. The left and the right columns correspond to the same experiment with and without the Hall effect, respectively.
As can be seen, in all the cases, virtually all of the injected slow wave at the base ($\beta \gg 1$) is stored as pure magnetic energy. The downgoing (negative) acoustic energy is supplied by mode conversion to the fast wave near the equipartition level.
The \alf{} wave is non-existent below the Hall window, i.e., $z < -0.23\,\mathrm{Mm}$. The solid horizontal line represents the total upward energy flux, which should remain constant everywhere as required by conservation of energy. Thus, it gives a measure of numerical errors involved in each case.
The dashed vertical line flags the location of $z_{\rm eq}$.

In the right column, $\bs{\mathcal{F}}_{\rm A}$ is identically zero everywhere, as expected. Notice the heavily reflected slow and converted/reflected fast waves at $\theta = 0$, as implicated by the amplitudes of $\bs{\mathcal{F}}_{\rm ac}$ and $\bs{\mathcal{F}}_{\rm mag}$ near the base. The small magnitude of the net upgoing flux ($\approx 4\%$) suggests that nearly $96\%$ of the original slow wave was reflected/mode converted into downgoing slow and fast waves. On the other hand, the small positive amplitude of $\bs{\mathcal{F}}_{\rm mag}$ at the base indicates that about $92\%$ of the input slow wave has been reflected as downgoing slow wave, and hence leaving about $4\%$ for the downgoing mode converted fast wave. Additionally, there is almost no reflection/conversion taking place at $\theta = 60^\circ$, which is suggestive that all of the injected slow wave has been transmitted through the barrier of the $\beta = 1$ layer.

In the left column, the Hall    conversion  strength at $\theta = 0$ is precisely the right amount for a perfect slow-\alf{} transformation. This is a critical amount beyond which $\mathcal{F}_{\rm A}^{\uparrow}$ begins to oscillate (see the plots for $\theta = 30^\circ$ and $60^\circ$).
Remember that the 2.5D configuration gains full 3D generality at $\theta = 0$. That is, despite the 2.5D nature of our equations, a full 3D approach would still yield identical results in the case of the vertical magnetic field. It is noteworthy that even though at $\theta = 0$ the high-$\beta$ slow wave in a vertically stratified medium becomes asymptotically incompressible, it is still intrinsically distinct from the \alf{} wave owing to their unique polarisation vectors.   

Furthermore, contrary to the non-Hall case and as a direct byproduct of the efficient slow-\alf{} conversion mediated by the Hall effect, no one of the waves is reflected back regardless of the angle $\theta$. This is verified by looking at the zero tails of $\bs{\mathcal{F}}_{\rm ac}$ and $\bs{\mathcal{F}}_{\rm A}$ at the bottom wall. At $\theta = 30^\circ$, $\mathcal{F}_{\rm A}^{\uparrow}$ whittles down as the Hall    conversion  strength ratchets up in comparison with that of the vertical field. Lastly, the case of $\theta = 60^\circ$ yields even less effectiveness in view of the much stronger    conversion  effect that spawns the cyclic pattern that culminates where the local slow energy flux is larger than that of the \alf{} wave. 
Thus, we observe that
\begin{enumerate}
    \item The strength of the slow-\alf{} Hall    conversion  per unit length is enhanced by widening of the angle $\theta$. 
    \item The length of the magnetic field lines in the main box is increased with $\theta$ by a factor of $\sec{\theta}$. As a result, field-guided waves such as the \alf{} and the high-$\beta$ slow waves remain    transformable for longer windows.
    \item The effectiveness can significantly decline or increase with heightening of the    conversion  effect. 
\end{enumerate}
As an important distinction, notice that the second observation above cannot apply to an input high-$\beta$ fast wave by virtue of its acoustic (or gravity, depending on $f$ and $k_x$) characteristics. We will elaborate on this in Section~\ref{subsec:twostage}.

For the remainder of this subsection, we shall progress by pairing up the parameters into three sub-spaces and survey the energy flux functions at the boundaries of the main box in each case.
\subsubsection{Survey parameters: $f-k_x$}\label{subsubsec:fkx}

The choice of $k_x$ and $f$ reflects the expected photospheric length and time scales of the wave sources. For granules, with typical extent 1~Mm and timescale of a few minutes, we might focus on $k_x=\mathcal{O}(1)$ $\rm Mm^{-1}$ and $f$ of a few mHz. For weak quiet-Sun magnetic field strengths of $B_0=20$ G say, the Alfv\'en speed at $z=-0.2$ Mm is only around 80 $\rm m\,s^{-1}$ and for $f=12$ mHz, as in Fig.~\ref{fig:dispersion}, we have $k_z\cos\theta\approx\omega/a\approx 945$ $\rm Mm^{-1}$, so excited slow waves with $k_x=\mathcal{O}(1)$ would be essentially vertical ($k_z\gg k_x$) at the base. This is consistent with the rapid divergence of the slow and Alfv\'en loci in Fig.~\ref{fig:dispersion} as $z$ diminishes below $z_{\rm eq}$.

However, $k_x$ becomes important at higher altitudes where $k_z$ has become comparable to $k_x$. In the figure, if $k_x$ were identically zero, the loci would be up-down symmetric, as would also be the case for vertical magnetic field $\theta=0^\circ$. The main effect of the non-zero $k_x$ is on the proximity of the Alfv\'en and fast loci. Although this has no practical consequences in 2D, because there is no available    conversion  mechanism, it does lead to powerful Alfv\'en-slow conversion in 3D, where wave vector, gravity and magnetic field direction are not co-planar \citep{CalHan11aa}. In that case, a dispersion diagram similar to Fig.~\ref{fig:dispersion} (not shown) indicates that an upgoing fast wave would    mode convert predominantly to the upgoing Alfv\'en wave. But if the figure were flipped ($k_x$ or $\theta$ negative), it would be the downgoing fast wave, post-reflection, that predominantly    transforms to the (now downgoing) Alfv\'en wave.

On the other hand, the slow-Alfv\'en    conversion  all occurs where $k_z$ is large, and therefore the effect of $k_x$ is minor. This will be seen in our computational results, where we observe near-symmetry in the Hall-mediated top escaping Alfv\'en flux about $\theta=0^\circ$.

\begin{figure}
    \centering
    \includegraphics[width=0.48\textwidth]{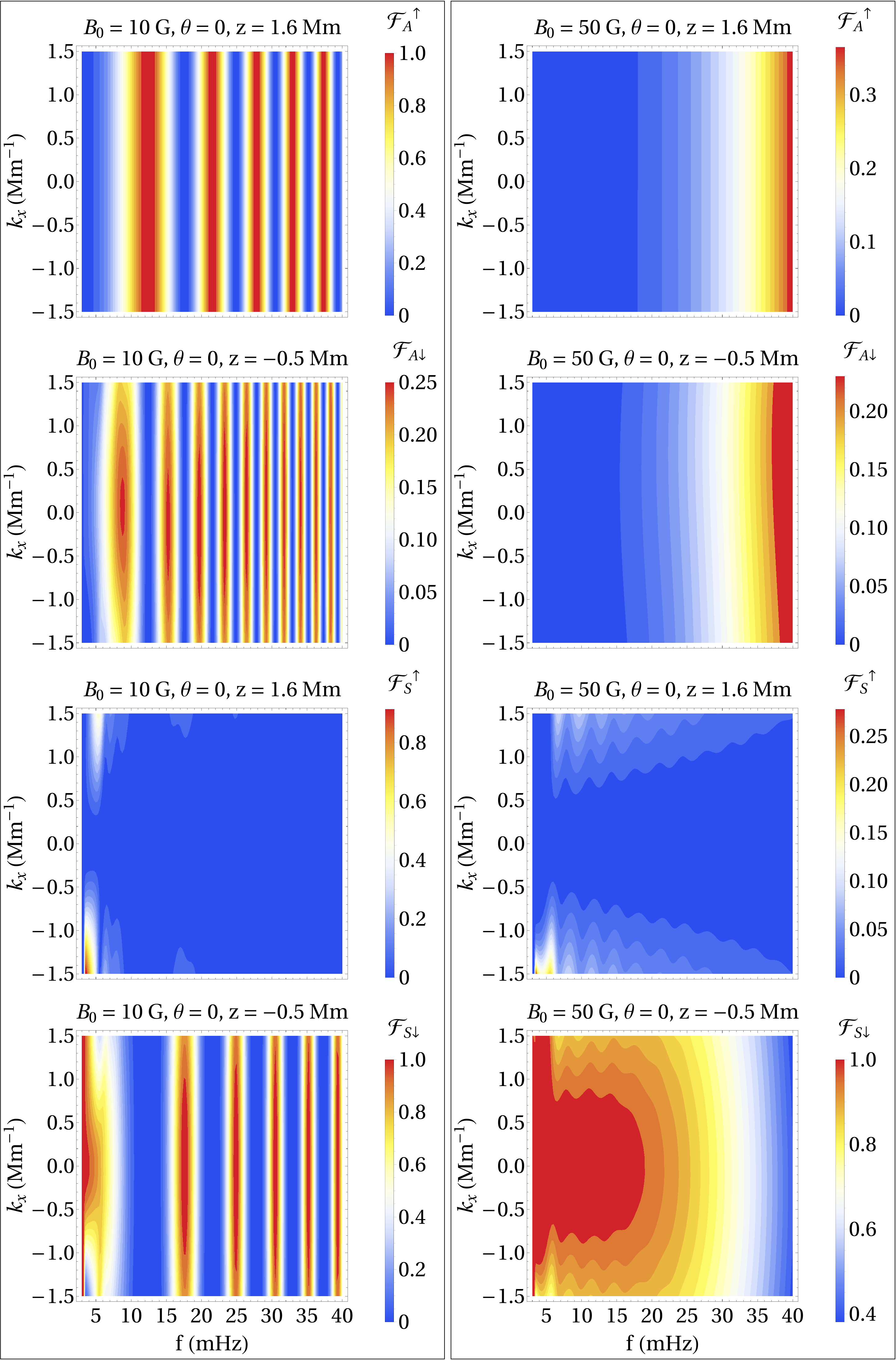}
    \caption{Outgoing slow and \alf{} wave-energy flux profiles at the boundaries as functions of the horizontal wave vector $k_x$ and the frequency $f$. Note the different colour-scaling of each panel. The plots correspond to $\theta = 0$ and two representative magnetic strengths  $10\,\mathrm{G}$ (left column) and $50\,\mathrm{G}$ (right column), as labelled. The upgoing \alf{} energy flux at the top manifests almost no variation across $-1.5\,\mathrm{Mm}^{-1} \le k_x \le 1.5\,\mathrm{Mm}^{-1}$ at either of the field intensities. This is because the vertical wave vector $k_z$ of the mode converted \alf{} wave is much greater than the $k_x$ of the source wave in the slow-Alfv\'en interaction region.}
    \label{fig:nuk}
\end{figure}

\begin{figure}
    \centering
    \includegraphics[width=0.48\textwidth]{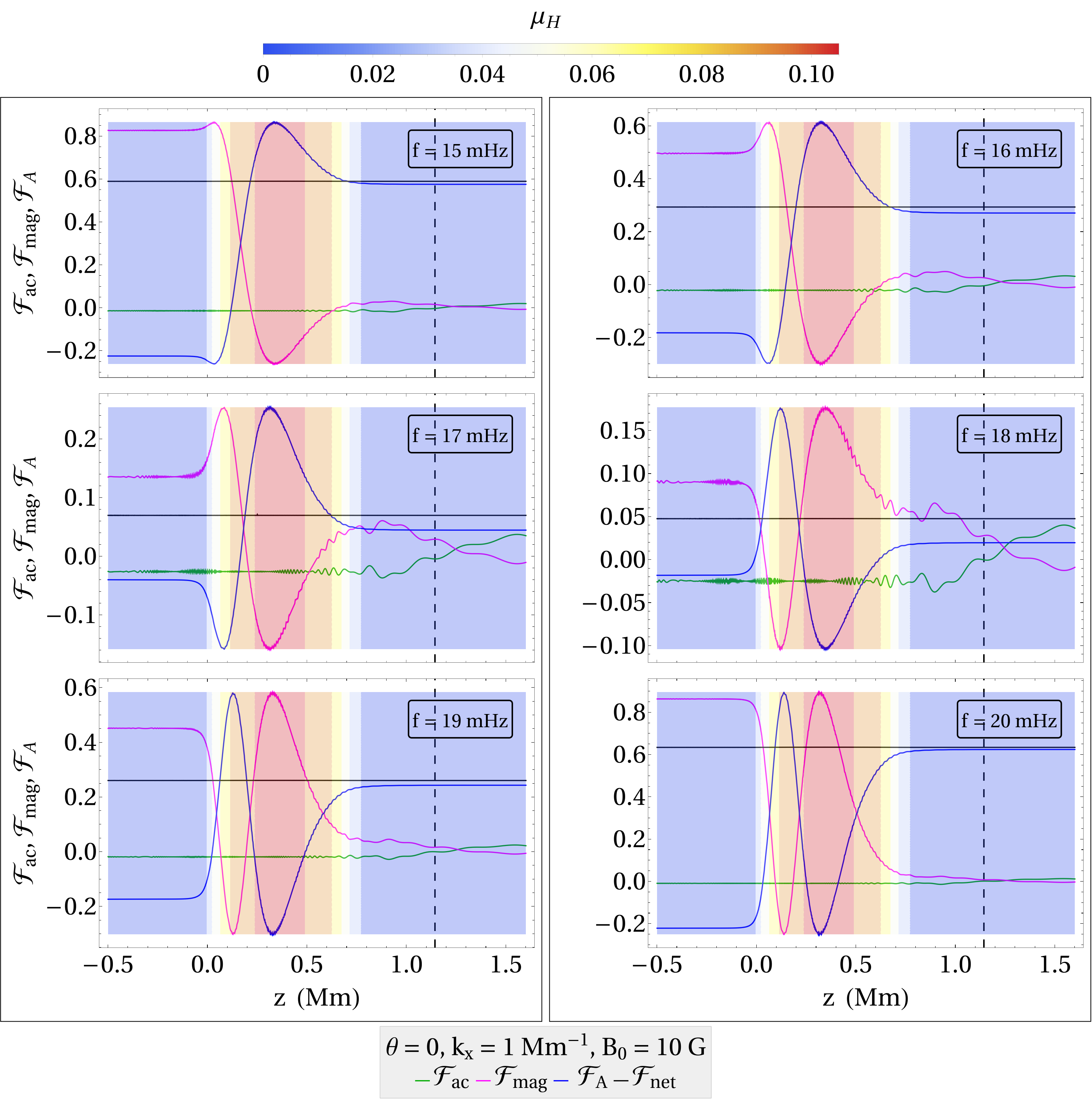}
    \caption{The energy flux functions associated with representative input slow waves at six frequencies from $15$ to $20\,\mathrm{mHz}$, as labelled. The magnetic field parameters are fixed to $B_0 = 10\,\mathrm{G}$ and $\theta = 0$ across the board. The strength of the Hall    conversion  is consistently augmented by increasing $f$. The effectiveness, however, oscillates with $f$, as the flux curves get modified with the variable    conversion  strength while the behaviour of the Hall-mediated region remains unchanged.}
    \label{fig:fluxes-freq}
\end{figure}

Figure~\ref{fig:nuk} displays the contour plots of outgoing slow and Hall-induced mode converted \alf{} energy fluxes as functions of $f$ and $k_x$ at two nominated magnetic intensities $10\,\mathrm{G}$ (left column) and $50\,\mathrm{G}$ (right column). The fast wave-energy flux is left out as it is zero for $z > z_{\rm eq}$ and is deducible by equation~(\ref{eq:enecon}) for $z < z_{\rm eq}$.
As seen in the first row, the effectiveness ($\mathcal{F}_{\rm A}^{\uparrow}$)
\begin{enumerate}
    \item Is very weakly dependent on $k_x$ for $\abs{k_x} \leq 1.5\,\rm{Mm}^{-1}$;
    \item Oscillates with $f$;
    \item Is remarkably scaled down with the larger $B_0$, requiring frequencies higher than $35$ mHz to achieve any significant upward \alf{} flux.
\end{enumerate}

The weak sensitivity of the effectiveness to $k_x$ at these frequencies results from the much larger vertical wavenumbers $k_z$ of the waves outweighing the effect of $k_x$ in the slow-Alfv\'en interaction region (where the loci are close in the dispersion diagram).

Increasing magnetic field strength $B_0$ reduces the    conversion  because of the Hall effect's inverse dependence on the ion cyclotron frequency $\Omega_{\rm i}=Z e B_0/m_{\rm i}$ through the dimensionless `Hall parameter' $\epsilon_H$ introduced in Sec.~\ref{sec:gov} \citep[see the discussion in Sec.~5 of][]{CalKho15aa}.

Moreover, the alternating patterns in $\mathcal{F}_{\rm A}^{\uparrow}$ suggest a periodic dependence of the effectiveness on some super-linear function of frequency. These patterns repeat at the higher magnetic strength in the right column, though widened and shifted upwards.

Unlike $\mathcal{F}_{\rm A}^{\uparrow}$, the behaviours of all the other fluxes are impacted by $k_x$. However, the influence of $B_0$ remains similar. 

As demonstrated by Figure~\ref{fig:fluxes-freq}, the periodicity of $\mathcal{F}_{\rm A}^{\uparrow}(f)$ originates in the ever-increasing strength of the Hall    conversion   with frequency via $\epsilon_H$ and size-invariability of the Hall domain. That is, as the    transformation intensifies with $f$ (signified by more dense cyclic patterns), the position of the intersection point of $\mathcal{F}_A$ and the upper boundary of the Hall domain traces out a periodic curve.

\subsubsection{Survey parameters: $f-\theta$}
\begin{figure}
    \centering
    \includegraphics[width=0.48\textwidth]{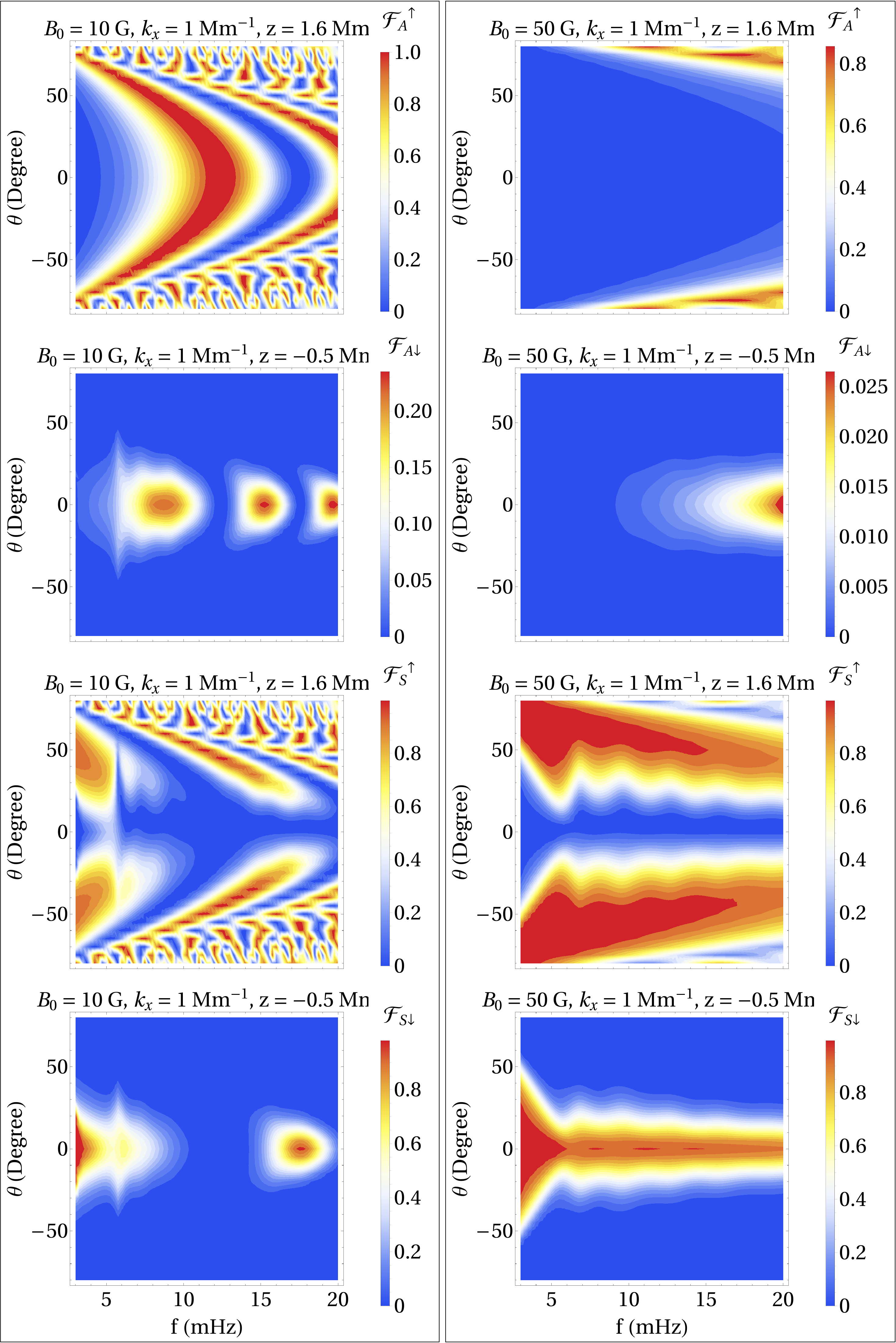}
    \caption{Outgoing energy fluxes at the boundaries carried by slow and \alf{} waves as functions of $\theta$ and $f$ at two fixed field intensities $B_0$ = 10 G (left column) and $B_0$ = 50 G (right column). The horizontal wavenumber $k_x=1$ $\rm Mm^{-1}$ is used throughout.
    The boomerang-shaped red band in the top panel flags the critical Hall    conversion  strength that is is exactly the right amount for perfect once-off conversions. The saw-tooth patterns signify extreme Hall    conversion  strengths wherein slight departures in $f$ or $\theta$ yield significant variations in the effectiveness.}
    \label{fig:nutheta}
\end{figure}

As discussed earlier, an increase in $\theta$ influences the effectiveness in two ways: by 1) directly bolstering the Hall    conversion  strength, and 2) lengthening the segment of the field lines lying within the weakly ionised portion of the atmosphere, and hence prolonging the    conversion  window for field-guided waves.
With that reminder, Figure~\ref{fig:nutheta} is presented to analyse the energy fluxes across the parameter sub-space $f$--$\,\theta$ at two representative magnetic intensities $10\,\mathrm{G}$ (left column) and $50\,\mathrm{G}$ (right column). 

In the left column, once-off slow to \alf{} mode conversions (similar to the top-left panel of Figure~\ref{fig:fluxes}) of varied effectiveness take place over the region bounded by $3\,\mathrm{mHz} \leq f \lesssim 13\,\mathrm{mHz}$ and $-50^\circ \lesssim \theta \lesssim 50^\circ$ (i.e., up to the first boomerang-shaped red stripe in $\mathcal{F}_{\rm A}^{\uparrow}$). Beyond this domain, conversions are of the cyclic form (similar to mid- and bottom-left panels of Figure~\ref{fig:fluxes}). The saw-tooth pattern signifies the areas of very high Hall    conversion  strengths, and hence a small displacement in either of the parameters have intense consequences on the effectiveness, giving rise to the highly fluctuating behaviour.
Notice that at low frequencies, this behaviour occurs only at very acute angles. Moreover, the $\mathcal{F}_{\rm A\downarrow}$ and $\mathcal{F}_{\rm S\downarrow}$ profiles reveal that it is the reflected slow wave passing through the Hall window that leads to the generation of downgoing \alf{} flux.  

At $B_0 = 50\,\mathrm{G}$, shown on the right column, identical patterns re-emerge, which once again highlight the role of $B_0$ as a scalar.

\subsubsection{Survey parameters: $f-B_0$}
\begin{figure}
    \centering
    \includegraphics[width=0.48\textwidth]{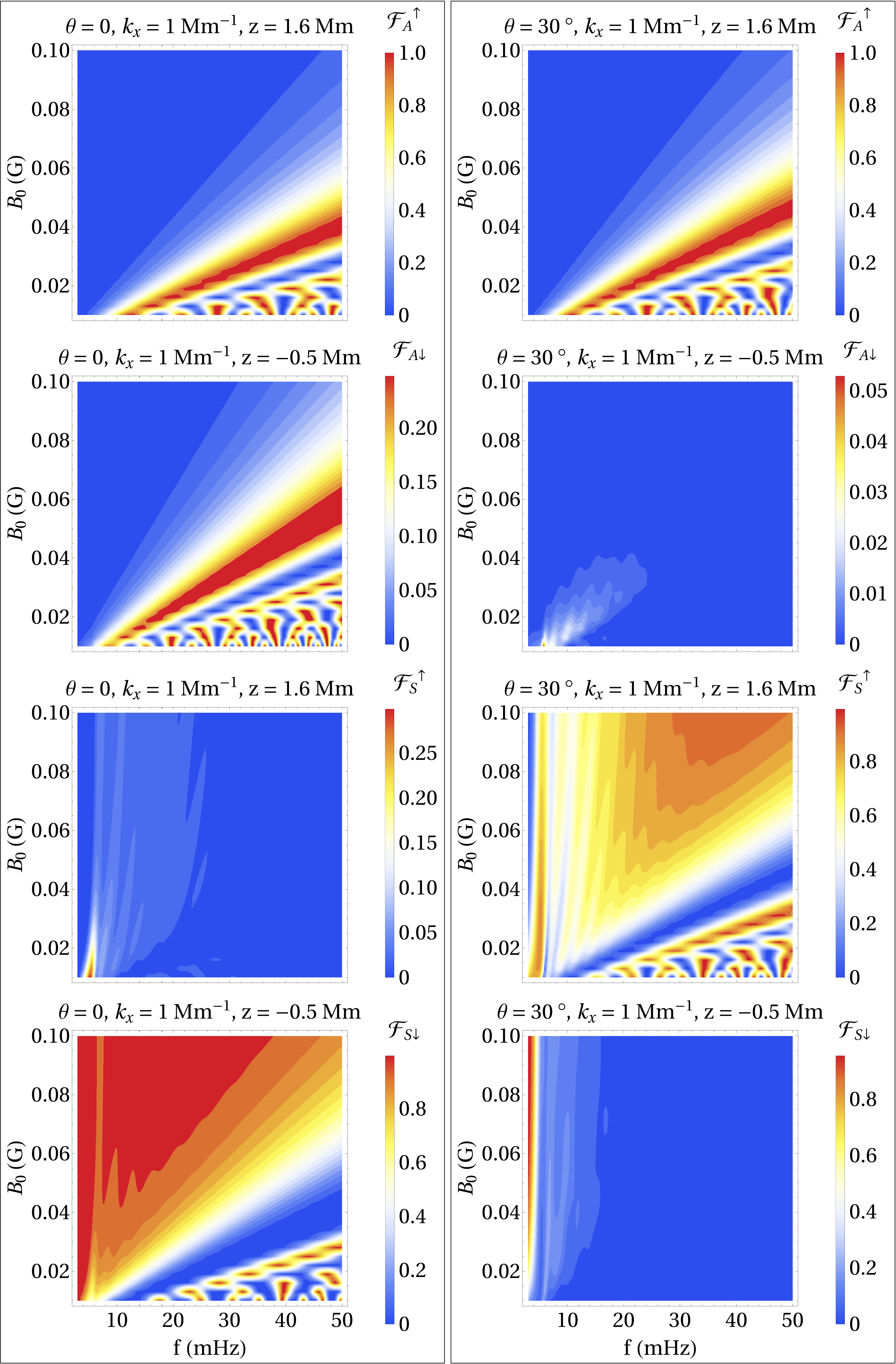}
    \caption{\alf{} and slow wave energy fluxes as functions of the magnetic strength $B_0$ and frequency $f$ passing through the boundaries of the main box at two representative field inclinations as labelled. The Hall effect is in operation. The upgoing \alf{} energy flux at the top remains almost insensitive to the field inclination, while the downgoing \alf{} flux at the bottom is significantly reduced with increasing angle. Conversely, the slow wave is heavily affected by the angle $\theta$, with the upgoing slow wave-energy intensified at the top and diminished at the bottom at $\theta = 30$ \textdegree.}
    \label{fig:nub0}
\end{figure}
The last parameter sub-space we wish to look at is $(f, B_0)$. Figure~\ref{fig:nub0} depicts the familiar four flux functions measured at the walls of the main box across the parameter domain $3\,\mathrm{mHz} \leq f \leq 50\,\mathrm{mHz}$ and  $10\,\mathrm{G} \leq B_0 \leq 100\,\mathrm{G}$. The solid oblique triangular red band signifies once-off slow-\alf{} conversions of 100\% effectiveness (i.e., conversions characterised with one whole revolution of the joint inter-modal polarisation vector). Above this band, the effectiveness declines as a direct result of weakening of the Hall    conversion  strength with $B_0$. Below this band, saw-tooth patterns identical to those of Figure~\ref{fig:nutheta} appear. These are spawned by the powerful Hall    conversion  effect stemmed from comparatively higher frequencies and lower magnetic intensities (i.e., conversions characterised with multiple revolutions of the joint inter-modal polarisation vector).

\subsection{Two-stage Fast to \alf{} Hall Conversion}\label{subsec:twostage}

Hall    transformation of magneto\-acoustic and Alfv\'en waves is a conservative process of polarisation rotation. As discussed in Sec.~\ref{subsec:slow2Alf}, in a high-$\beta$ plasma it directly affects the slow MHD wave, which is magnetically dominated and asymptotically transverse to $\bs{k}$ and to $\bs{k}\vcross\bs{B}_0$ as $\beta\to\infty$.

On the other hand, in a low-$\beta$ plasma, it is the fast wave that is magnetically dominated and asymptotically transverse to $\bs{B}_0$ (as $\beta\to0$). The role and efficacy of the Hall effect in this domain was explored by \cite{CalKho15aa} and confirmed computationally by \cite{Gonzalez-Morales2019} at $B_0=500$ G, though frequencies of order 1 Hz were then found necessary to produce a significant effect. This may be understood in terms of the dimensionless Hall parameter $\epsilon_H=\omega/f_{\rm i}\Omega_{\rm i}$ introduced in Sec.~\ref{subsubsec:fkx}. Only as $\epsilon_H$ approaches order unity is there substantial rotation.

With this in mind, we now explore the two-stage process by which fast waves launched from the base convert to Alfv\'en waves via the Hall effect, as distinct from the 3D geometric process of \cite{CalHan11aa}. To do this, the predominantly acoustic fast wave in $a<c_s$ partially transmits as an acoustic wave (now slow) in $a>c_s$ through the equipartition layer $z_{\rm eq}$, and partially converts to a magnetically dominated (now fast) wave. Provided this fast wave is still in the Hall window, it may be rotated and Alfv\'en waves produced. But for this to occur, $z_{\rm eq}$ must be lowered substantially compared to the cases of Figs.~\ref{fig:fluxes} and \ref{fig:fluxes-freq} to place it in the Hall region, which requires stronger magnetic fields. But then, higher frequencies will be required.

In this subsection, we first confirm that no significant Alfv\'en flux arises from fast waves at low field strengths and low frequencies, but then explore field strengths of 100 to 400 G characteristic of weak plage and frequencies of several hundred mHz.

\begin{figure}
    \centering
    \includegraphics[width=0.48\textwidth]{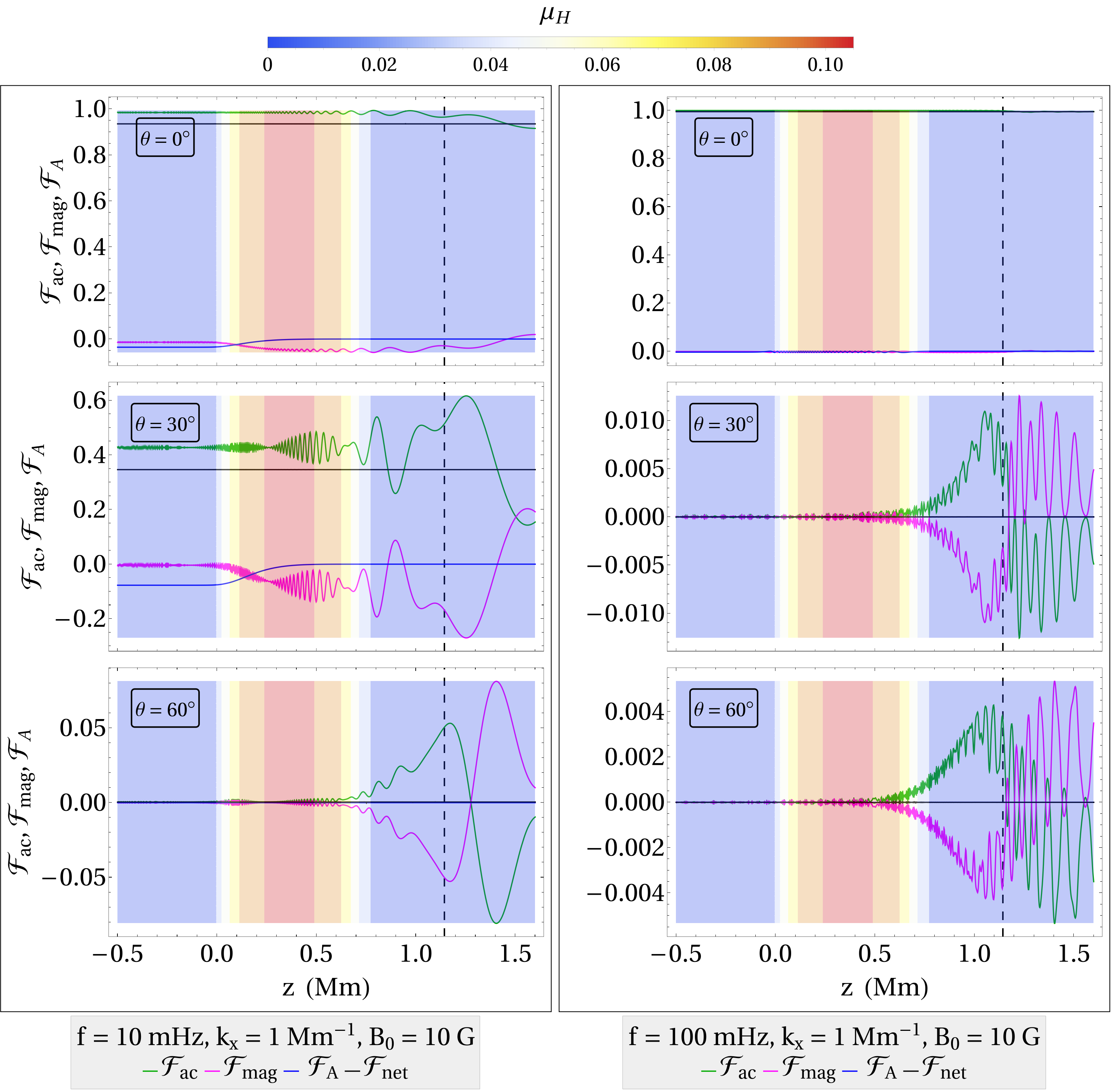}
    \caption{The energy flux functions corresponding to two injected fast waves at $10\,\mathrm{mHz}$ (left column) and $100\,\mathrm{mHz}$ (right column). Note that all of the injected energy in subsurface regions near the bottom boundary is now stored in $\bs{\mathcal{F}}_{\rm ac}$, duly indicating that the fast wave is predominantly of acoustic nature there. The solid horizontal and the dashed vertical lines represent the net upward energy flux and the position of the equipartition level, respectively. The colour-shading exhibits the profile of $\mu_H$ across the main box. Notice the zero tail of $\bs{\mathcal{F}}_{\rm A}$ near the top boundary in all the cases, which is an indicator that none of the fast wave was directly converted into upgoing \alf{} wave. Moreover, the zero tail of $\bs{\mathcal{F}}_{\rm mag}$ at the base suggests that none of the wave returns as slow mode. However, the small non zero tails of $\bs{\mathcal{F}}_{\rm A}$ at $f = 10\,\mathrm{mHz}$ and $\theta = 0, 30$ degree indicate a two-stage conversion into downward \alf{} wave: First, a portion of the injected fast wave is reflected as downgoing slow wave. Second, the reflected slow wave    converts to the downgoing \alf{} in the Hall dominated region and reaches the base as pure \alf{} wave.}
    \label{fig:fastfluxes}
\end{figure}

\begin{figure}
    \centering
    \includegraphics[width=0.48\textwidth]{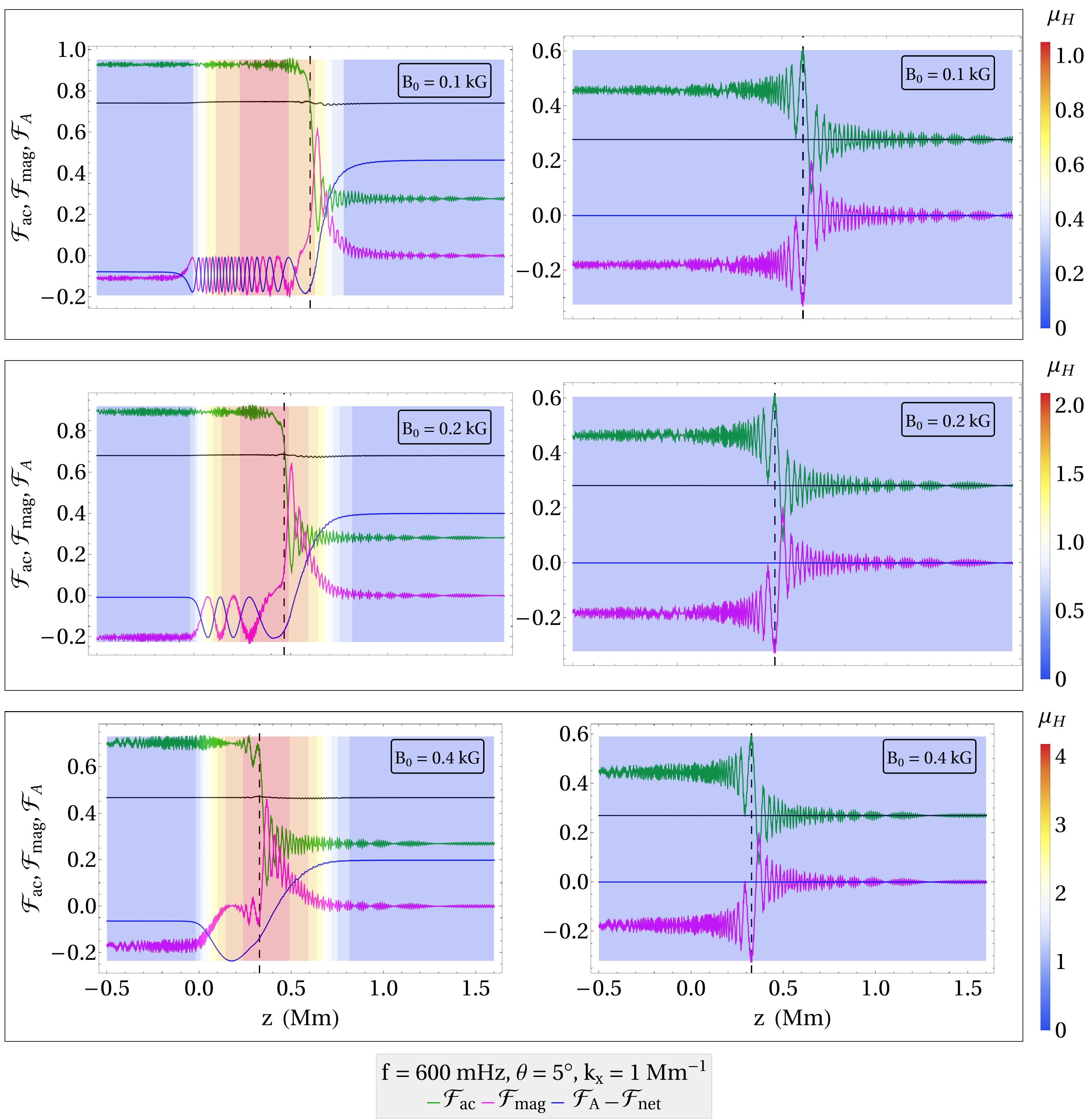}
    \caption{The behaviours of the energy flux functions of an input fast wave under three different magnetic intensities, with (left column) and without (right column) the Hall effect. The magnetic intensities are chosen such that $z_\mathrm{eq}$ is contained within the Hall window. The two-stage fast to \alf{} Hall-induced conversion is clearly at work: 
    first, the input fast wave undergoes a non-Hall conversion to slow wave at $z_{\mathrm{eq}}$. Imposed by the Hall effect, The resultant slow wave then faces another conversion to \alf{} wave.  
    Notice the drop in the amount of \alf{} flux with increasing magnetic intensity, despite the lower locations of $z_{\mathrm{eq}}$ corresponding to higher $B_0$. Therefore, the power of the slow-\alf{} Hall    conversion  set by the magnetic intensity plays a more important role than does the length of the distance through the Hall window over which the waves are allowed to exchange energy.}
    \label{fig:fastHighBeta}
\end{figure}

Let us first consider the case where $z_{\rm eq}$ lies above the Hall domain. In this case, any upgoing \alf{} flux can only result from a direct Hall-mediated fast-\alf{} conversion. 
Figure~\ref{fig:fastfluxes} presents the energy flux diagrams of such a setup for two normalised injected fast waves at 10~mHz (left column) and 100~mHz (right column) with $B_0 = 10$~G and $k_x = 1\,\rm{Mm}^{-1}$.
Evidently, $\mathcal{F}_{\rm A}^{\uparrow} = 0$ in all the cases, but there are smatterings of downgoing \alf{} flux at $f = 10$ mHz and $\theta = 0, 30^\circ$. Of course, these are produced out of the small amounts of reflected/converted downgoing slow waves passing through the Hall window. 

To interpret these results, we may first investigate the prospect of fast-slow conversions for such wave characteristics with reference to the acoustic-gravity propagation diagram of Figure~\ref{fig:regions}. Given $k_x = 1$ $\rm{Mm}^{-1}$, the waves in the left column are predominantly (obliquely propagating) gravity modes, while the right column consists of essentially (vertically propagating) acoustic modes.  In the special case of $f = 100$ $\rm{mHz}$ and $\theta = 0$, we find $\bs{B}_0 \parallel \bs{k}$ (due to disproportionately large $k_z$), and hence no fast-slow conversion is possible. On the other hand, despite the potential for fast-slow conversion in all the other cases, little to no conversion is witnessed on account of weak geometrical coupling effects at these frequencies. 

Let us now modify the field strength such that $z_{\rm eq}$ is re-located inside the Hall window, and amplify the frequency to guarantee stronger fast-slow    transformations. Figure~\ref{fig:fastHighBeta} shows the flux functions of an input fast wave at $f = 600\,\rm{mHz}$ and three nominated magnetic field strengths, with the Hall term switched on (left columns) and off (right column). As evident in the right column, all fast-slow conversions in the absence of the Hall effect are only weakly sensitive to the location of $z_{\rm eq}$ at these field strengths. Notice that due to very large values of $k_z$ at such high frequencies, $\bs{k} \sim k_z \hat{\bs{k}}$. Moreover, the acoustic energy carried by the input fast wave undergoes substantial transformation at $z_{\rm eq}$ and splits into magnetic and acoustic parts. Without the Hall effect, all the resultant magnetic energy is reflected back as (magnetic) slow waves.

Upon introducing the Hall effect (left column), the resultant magnetic energy derived from fast-slow conversions can now undergo a second transformation into \alf{} waves through the Hall effect. The mildly jagged patterns in the net upward energy flux (solid horizontal line) indicate minor numerical errors due to highly oscillatory solutions.
Notice the larger net upward fluxes compared with those of the right column. Furthermore, notice the relatively larger leaps in $\mathcal{F}_{\rm mag}$ at $z_{\rm eq}$. This is due to the fact that magneto\-acoustic and \alf{} waves in 2.5D are geometrically decoupled, and hence $\mathcal{F}_{\rm A}$ is invisible to geometric coupling effects. Thus, as acoustic energy transforms into the \alf{} type via the Hall effect, it depletes the available magnetic energy budget, which is then compensated by further geometry-induced acoustic-magnetic conversion.  
Finally, looking at the position of $z_{\rm eq}$ set by $B_0$, it is clear that the strength of the Hall    conversion  depends much more strongly on $B_0$ itself, rather than $z_{\rm eq}$. That is, even though $z_{\rm eq}$ sits much lower in the atmosphere at $B_0 = 400\,\rm{G}$, nevertheless such a large field intensity severely diminishes the strength of the Hall    conversion   to the point that the net effect is the relatively smaller effectiveness in contrast with the lower $B_0$ cases.

\begin{figure}
    \centering
    \includegraphics[width=0.48\textwidth]{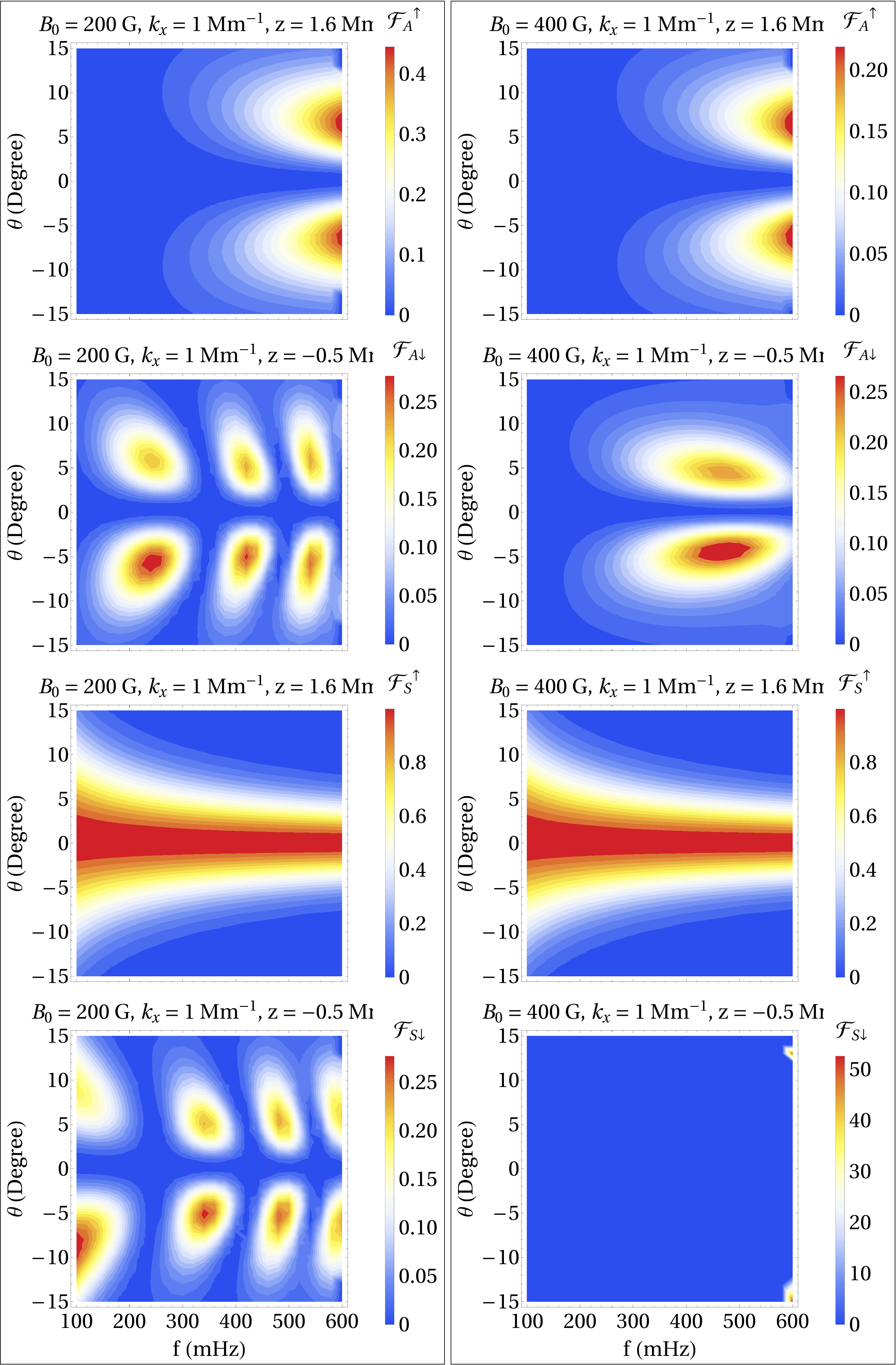}
    \caption{Contour maps of mode converted energy fluxes corresponding to high-$\beta$ injected fast waves at various frequencies and field inclinations and two representative strong field intensities 200 G (left) and 400 G (right). For both choices of $B_0$, the equipartition layer sits inside the Hall window, giving rise to the two-stage fast to Alfv\'en mode conversion.
    Notice the $\theta = 0^\circ$ band which yields no conversion to Alfv\'en waves.}
    \label{fig:fasthighB}
\end{figure}

To close this section, let us investigate the effectiveness of two-stage fast-\alf{} conversion in the $\theta-f$ parameter subspace. 
As a point of caution, substantial fast-slow conversions require high frequencies ($\sim$ Hz) which entail highly oscillatory behaviours and pose numerical problems. Due to these numerical challenges, we were only able to cover frequencies up to $600\,\rm{mHz}$ and magnetic inclinations no more than $15^\circ$. However, revealing aspects are still present at these ranges. Figure~\ref{fig:fasthighB} depicts the contour maps of the slow and \alf{} energy fluxes at $z_{\rm top}$ and $z_{\rm bot}$ for two representative field intensities $B_0 = 200\,\rm{G}$ and $B_0 = 400\,\rm{G}$. 
The bands of zero flux associated with $\theta \sim 0$ are due to vertical propagation of the input acoustic fast waves handing all the energy directly to acoustic slow waves at the top. Considerable amounts of upgoing \alf{} waves begin to appear for $\theta \gtrsim \abs{2^\circ}$ and $f \gtrsim 500\,\rm{mHz}$. The effectiveness maximises for $4^\circ \lesssim \theta \lesssim 10^\circ$, and analogous to the slow-\alf{} conversion, favours higher frequencies. However, the crucial difference here is that not only do high frequencies boost the Hall    conversion  effect, but they are required to ensure significant fast-slow conversions if \alf{} waves are to be generated.

\section{Conclusion}\label{sec:summary}
Dispersion diagrams such as Figure~\ref{fig:dispersion} provide useful insights into where MHD wave modes may interact, either because their loci are very close over an extended interval, such as for slow and Alfv\'en curves in $a\ll c_s$ and fast and Alfv\'en in $a> c_s$, or because of avoided crossings, seen for slow and fast curves near $a=c_s$. In the case of the fast-slow avoided crossings, the closeness of the gap indicates the strength of the    convertibility, which operates in ideal MHD in both 2D and 3D.

However, the other two opportunities are only realised if a    conversion  mechanism is present, in addition to phase space proximity. In the case of fast-Alfv\'en this is afforded by three-dimensionality, where $\bs{k}$, $\bs{g}$ and $\bs{B}_0$ are non-coplanar \citep{CalHan11aa}, but both slow-Alfv\'en (in $a\ll c_s$) and fast-Alfv\'en (in $a> c_s$) are also potentially mediated by the Hall effect, provided the Hall parameter $\epsilon_H$ is of order unity in the relevant regions.

Even at low frequencies ($\sim 10$ mHz) and low magnetic field strengths ($\sim10$ G), the Hall effect is very potent in    mode converting slow and Alfv\'en waves at low altitudes (see Figs.~\ref{fig:fluxes} and \ref{fig:fluxes-freq}), and may in fact produce several oscillations from fast to \alf{} and back again, with ultimate net effect depending on how many such oscillations fit within the Hall window.

Fast waves in $z<z_{\rm eq}$ are not    transformable to Alfv\'en waves, since their phase space loci are well-separated and because their predominantly acoustic natures are not susceptible to the Hall effect. However, on (ideal) mode conversion to now magnetically dominated and predominantly $\bs{B}_0$-transverse fast waves in $z>z_{\rm eq}$, they are then prone to Hall conversion to Alfv\'en slightly below the fast wave reflection height. This only occurs to appreciable extent though if the interaction is in the Hall window, which in turn can only happen if $z_{\rm eq}$ is low enough. This requires stronger magnetic fields (hundreds of gauss), and therefore frequencies of hundreds of mHz, since $\epsilon_H\propto \omega/B_0$.

For the various scenarios considered, dispersion diagrams (Fig.~\ref{fig:dispersion}), flux vs.~height plots (Figs.~\ref{fig:fluxes}, \ref{fig:fluxes-freq}, \ref{fig:fastfluxes} and \ref{fig:fastHighBeta}), and several two-parameter net flux contour plots (Figs.~\ref{fig:nuk} [$f$ and $k_x$], 
\ref{fig:nutheta} [$f$ and $\theta$], \ref{fig:nub0} [$f$ and $B_0$] and \ref{fig:fasthighB} [$f$ and $\theta$ for high $B_0$]), reveal a rich picture of Hall conversion to Alfv\'en waves, with implications for upper atmospheric heating and interpretation of Alfv\'enic wave observations.

One shortcoming of our analysis is that it is only 2.5D, having assumed that $\bs{k}$ is in the vertical ($x$--$z$) plane of the magnetic field lines. It therefore does not address geometrical coupling between slow and Alfv\'en, which will be considered elsewhere. However, this is irrelevant in the important case of vertical magnetic field $\theta=0^\circ$, since the choice of $x$-direction is now arbitrary. In that case, our various diagrams indicate that strong direct slow-to-Alfv\'en    conversion  at low altitudes is unavoidable at low field strengths. 

Our survey tells us that in regions of weak magnetic field, both
\begin{enumerate}
    \item Slow waves generated by convective motions near the solar surface easily convert to Alfv\'en waves on their way upwards, perhaps many times over, especially at higher frequencies; and
    \item Alfv\'en waves generated by convective motions near the solar surface easily convert to slow waves on their way upwards, perhaps many times over, especially at higher frequencies.
\end{enumerate}
This operates even at low frequencies, and is amplified further by magnetic field inclination. Higher magnetic field strengths require proportionately higher frequencies for similar effects. The overall effect is a sort of shuffling process between slow and Alfv\'en, with only the latter able to contribute to the Alfv\'enic `dark energy' content of the corona \citep{McIde-Car11aa,McIDe-12aa}.



\section*{Data Availability}

The data underlying this article will be shared upon reasonable request to the corresponding author.



\bibliographystyle{mnras}
\bibliography{manuscript} 




\appendix

\section{Linearised MHD equations at the boundaries}\label{sec:app1}
Here, we present the ODEs for the plasma displacement vector $\bxi$ drawn from equations~(\ref{eq:mhd}) at the   isothermal boundaries. 

As discussed earlier, magnetoacoustic-\alf{} coupling at the boundaries is entirely terminated due to vanishing of the Hall parameter in those heights (Figure~\ref{fig:muH}). Thus, setting $\mu_H = 0$ in equation~(\ref{eq:induction})   and substituting equations~(\ref{eq:mass} -- \ref{eq:energy}) into equation~(\ref{eq:momentum}), we obtain a vectorial ODE of 4-th order. Using the magnetic field coordinate system ($\hat{\bs{e}}_\parallel,\,\hat{\bs{e}}_\perp,\, \hat{\bs{e}}_y$) introduced in Section \ref{sec:BCs}, the individual components of this equation can then be sifted out according to,
\begin{subequations}
\begin{multline}\label{eq:xiperp}
    \left(4 (s^2 - \kappa^2 - 4 \kappa^4 - 4 \kappa^2 \kappa_{0}^2 + 4 (\kappa^4 + s^2 \kappa_{z}^2) \sec^{2}\theta \right. \\
\left. \hspace{2.95cm}- 4 i \kappa^3 \tan \theta 
+ s^2 \tan^{2}\theta\right) \xi_{\perp} \\
+ 4 s \left(2 \kappa^2 + \kappa_{0}^2 + (3 s^2 - \kappa^2) \sec^{2}\theta 
+ 4 i \kappa^3   \tan \theta \right) \xi_{\perp}'\\
+ 2 s^2 \left(1 + 2 \kappa_{0}^2 + 2 (s^2 - \kappa^2) \sec^{2}\theta 
- 4 i \kappa  \tan \theta \right) \xi_{\perp}'' \\
+ 4 s^3 (1 - i \kappa  \tan \theta )\xi_{\perp}^{(3)} 
+s^4 \xi_\perp^{(4)} = 0,
\end{multline}
\begin{multline}\label{eq:xipar}
     \xi_\parallel = \Delta^{-1}\biggl\{
    2 i \gamma ^2 \kappa  \nu^2 s^3 \cos
   ^2\theta\,  \xi_\perp^{(3)}\\
   +2 i \gamma  \nu^2 s^2 \cos\theta    \left[(3 \gamma -2) \kappa  \cos
   \theta -2 i \gamma  \sin\theta  
   \left(\kappa^2+\nu^2\right)\right]\xi_\perp''\\
   + i \gamma  \nu^2 s  \biggl[-2 i \gamma
    \sin 2 \theta  \left(\kappa^2+\nu
   ^2\right)-\kappa  \left(\gamma  \left(4
   \kappa^2-1\right)+2\right) \cos 2 \theta\\
   +\kappa  \left(-4 \gamma  \kappa
   ^2+\gamma +8 \gamma  s^2-2\right)\biggr]\xi_\perp'\\
  + 8 i\, \xi_\perp \biggl[2 \gamma  \kappa  \nu^2
   \cos^2\theta  \left(\kappa^2+(\gamma
   -1) s^2\right)\\
   +2 \sin\theta  
   \left(\gamma  \kappa  \nu^2 s^2 \sin
   \theta +i \cos\theta   \left(\gamma ^2
   \kappa^2 \nu^2 \left(\kappa^2+\nu
   ^2\right) \right. \right. \\
   \left. \left. +s^2 \left((\gamma -1) \kappa
   ^2-\gamma ^2 \nu
   ^4\right)\right)\right)\biggr]\biggr\},
\end{multline}
\begin{align} \label{eq:eta}
    s^2 \cos^2\theta\, \eta'' + s \cos\theta  (\cos\theta  &- 4 i \kappa  \sin\theta ) \eta' \nonumber \\ 
&+4 (s^2-\kappa^2 \sin^2 \theta) \eta = 0,
\end{align}
\end{subequations}
where   primes denote $s$-derivatives and
\begin{align*}
\Delta = 8 s^2 \biggl(\cos 2 \theta  \left(\gamma ^2
   \nu ^4-(\gamma -1) \kappa^2\right)-\gamma
   ^2 \nu ^4-i (\gamma -2) \gamma  \kappa 
   \nu^2 \sin 2 \theta \\
   +\kappa^2
   \left(\gamma  \left(2 \gamma  \nu
   ^2-1\right)+1\right)\biggr).
\end{align*}
 The various parameters used in the above equations are set out in equations~(\ref{eq:nu}) through to (\ref{eq:kapaz}). Note that although the 2.5D magnetoacoustic oscillations are jointly controlled by $\xi_\parallel$ and $\xi_\perp$, using the magnetic field coordinates allows for the independence of $\xi_\perp$ from $\xi_\parallel$ seen in equation~(\ref{eq:xiperp}).

\section{Hall coupled Linearised MHD equations in the main box}\label{sec:app2}
  Here, we lay out the set of second order ODEs in terms of $\xi_\perp$, $\xi_\parallel$, and $\eta$ to be solved numerically throughout the domain. 

To combine equations~(\ref{eq:mhd}) into a single vectorial ODE in the magnetic field coordinates, this time we start by eliminating $\bs{j}_1\vcross\bs{B}_0$ by substituting it from the momentum equation~(\ref{eq:momentum}) into the induction equation~(\ref{eq:induction}). The resultant $\bs{b}$ can then enter equation~(\ref{eq:amper}) to obtain $\bs{j}$ in terms of $\bs{\xi}$, $\rho_1$, $p_1$ and their derivatives. Finally, the new $\bs{j}$, along with $\rho_1$, $p_1$, and $\bs{\xi}$ from equations~(\ref{eq:mass}), (\ref{eq:energy}), and (\ref{eq:displacement}) are substituted back into equation~(\ref{eq:momentum}), which yields the $x$-component of the momentum equation
\begin{subequations}\label{eq:mainPDEs}
\begin{multline}
    \xi  \left(-a^2  k_x^2 \rho_0  \cos^{2}\theta -c_{\rm{s}}^{2}  k_x^2 \rho_0 +\omega^2  \rho_0 \right)\\
    +\zeta  \left(a^2  k_x^2 \rho_0 \sin\theta  \cos\theta  -i g k_x \rho_0 \right) \\
    -a^2 \rho_0  \zeta'' \sin\theta  \cos\theta   +a^2 \rho_0  \xi''  \cos^{2}\theta  +i c_{\rm{s}}^{2}  k_x \rho_0  \zeta' \\
    +\eta  \left(i \omega \, k_x^2 \mu_H \cos\theta  -i \omega \, \mu_H'' \cos\theta  \right) -2 i \omega  \eta'  \mu_H' \cos\theta  \\
    -i \omega  \mu_H  \eta'' \cos\theta  = 0 ,
\end{multline}
  the $y$-component
\begin{multline}
    \xi  \left(i H k_x \rho_0  \left(\left(c_{\rm{s}}^{2}\right)' +g\right)+a^2  \sin\theta  \cos\theta\,  H k_x^2 \rho_0 -i c_{\rm{s}}^{2}  k_x \rho_0 \right)\\
    +\zeta'  \left(H \rho_0  \left(c_{\rm{s}}^{2}\right)' -c_{\rm{s}}^{2}  \rho_0 \right)+\zeta''  \left(a^2  \sin^{2}\theta\,  H \rho_0 +c_{\rm{s}}^{2}  H \rho_0 \right)\\
    +H \zeta  \rho_0  \left(\omega^2 -a^2  \sin^{2}\theta\,  k_x^2\right)-a^2  \sin\theta  \cos\theta\,  H \rho_0  \xi'' \\
    +i c_{\rm{s}}^{2}  H k_x \rho_0  \xi' +\eta  \left(i \omega  \sin\theta\,  H \mu_H'' -i \omega  \sin\theta\,  H k_x^2 \mu_H \right)\\
    +2 i \omega  \sin\theta\,  H \eta'  \mu_H' +i \omega  \sin\theta\,  H \mu_H  \eta'' = 0,
\end{multline}
and   the $z$-component
\begin{multline}
    \omega  H^2 \left(\omega^2 -a^2  \sin^{2}\theta\,  k_x^2\right) \eta  \rho_{0}^3 +i \omega  a^2  H^2 \sin{2\theta}\,  k_x \eta'  \rho_{0}^3 \\
    +\omega  a^2  \cos^{2}\theta\,  H^2 \eta''  \rho_{0}^3 +i \omega^2  \cos\theta\,  H^2 \mu_H  \xi''  \rho_{0}^2 \\
    +\xi'  \left(g i \cos\theta\,  H^2 \rho_{0}^2  \mu_H  k_x^2+i c_{\rm{s}}^{2}  \cos\theta\,  H^2 \rho_0  \mu_H  \rho_{0}'  k_x^2 \right.\\
    \left. -i c_{\rm{s}}^{2}  \cos\theta\,  H^2 \rho_{0}^2  \mu_H'  k_x^2-\omega^2  H^2 \sin\theta\,  \rho_{0}^2  \mu_H  k_x \right.\\
    \hspace{2cm}\left. +2 i \omega^2  \cos\theta\,  H^2 \rho_{0}^2  \mu_H' \right)\\ 
    +\left(g \cos\theta\,  k_x \rho_{0}^2  \mu_H  H^2+c_{\rm{s}}^{2}  \cos\theta\,  k_x \rho_0  \mu_H  \rho_{0}'  H^2 \right.\\
    \hspace{2cm}\left. -c_{\rm{s}}^{2}  \cos\theta\,  k_x \rho_{0}^2  \mu_H'  H^2\right) \zeta'' \\ 
    +\zeta'  \left(-c_{\rm{s}}^{2}  \cos\theta\,  k_x \mu_H  \lrp{\rho_{0}'}^2 H^2+g i \sin\theta\,  k_x^2 \rho_{0}^2  \mu_H  H^2 \right.\\
    \hspace{0.7cm}\left. + \omega^2  \cos\theta\,  k_x \rho_{0}^2  \mu_H  H^2+i c_{\rm{s}}^{2}  \sin\theta\,  k_x^2 \rho_0  \mu_H  \rho_{0}'  H^2 \right.\\
    \hspace{0.7cm}\left. - i c_{\rm{s}}^{2}  \sin\theta\,  k_x^2 \rho_{0}^2  \mu_H'  H^2+2 g \cos\theta\,  k_x \rho_{0}^2  \mu_H'  H^2\right. \\
    \hspace{0.7cm}\left. - \cos\theta\,  k_x \rho_{0}^2  \left(c_{\rm{s}}^{2}\right)'  \mu_H'  H^2+c_{\rm{s}}^{2}  \cos\theta\,  k_x \rho_0  \rho_{0}'  \mu_H'  H^2 \right.\\ 
    \hspace{0.7cm}\left. - \cos\theta\,  k_x \rho_0  \mu_H  \left(\left(g-\left(c_{\rm{s}}^{2}\right)' \right) \rho_{0}' -c_{\rm{s}}^{2}  \rho_{0} '' \right) H^2\right.\\ 
    \hspace{0.7cm}\left. -c_{\rm{s}}^{2}  \cos\theta\,  k_x \rho_{0}^2  \mu_H''  H^2-g \cos\theta\,  k_x \rho_{0}^2  \mu_H  H\right) \nonumber
\end{multline}
\begin{multline}
    +\zeta  \left(g \cos\theta\,  k_x \mu_H  \lrp{\rho_{0}'}^2 H^2-g \cos\theta\,  k_x \rho_0  \rho_{0}'  \mu_H'  H^2\right.\\
    \hspace{0.7cm}\left.+i \sin\theta\,  k_x^2 \rho_0  \left(\rho_0  \left(\mu_H  \omega^2 +g \mu_H' \right)-g \mu_H  \rho_{0}' \right) H^2 \right.\\
    \hspace{0.7cm}\left. - g \cos\theta\,  k_x \rho_0  \mu_H  \rho_{0} ''  H^2+\cos\theta\,  k_x \rho_{0}^2  \left(\mu_H'  \omega^2 +g \mu_H'' \right) H^2 \right.\\ 
    \hspace{0.7cm}\left. - i g \sin\theta\,  k_x^2 \rho_{0}^2  \mu_H  H-g \cos\theta\,  k_x \rho_{0}^2  \mu_H'  H \right.\\
    \hspace{0.7cm}\left. +g \cos\theta\,  k_x \rho_{0}^2  \mu_H  H'(z)\right)
    \allowdisplaybreaks \\
    +\xi  \left(-H^2 \sin\theta\,  \rho_0  \left(c_{\rm{s}}^{2}  \mu_H  \rho_{0}' +\rho_0  \left(g \mu_H -c_{\rm{s}}^{2}  \mu_H' \right)\right) k_x^3 \right. \\ 
    \hspace{0.7cm}\left. - i c_{\rm{s}}^{2}  \cos\theta\,  H^2 \mu_H  \lrp{\rho_{0}'}^2 k_x^2\right.\\
    \hspace{0.7cm}\left. + i \cos\theta\,  H^2 \rho_0  \left(c_{\rm{s}}^{2}  \rho_{0}'  \mu_H' +\mu_H  \left(\left(c_{\rm{s}}^{2}\right)'  \rho_{0}' +c_{\rm{s}}^{2}  \rho_{0} '' \right)\right) k_x^2 \right.\\ 
    \hspace{0.7cm}\left. +i \cos\theta\,  H^2 \rho_{0}^2  \left(\left(g-\left(c_{\rm{s}}^{2}\right)' \right) \mu_H' -c_{\rm{s}}^{2}  \mu_H'' \right) k_x^2 \right.\\
    \hspace{0.7cm}\left. -\omega^2  H^2 \sin\theta\,  \rho_{0}^2  \mu_H'  k_x+i \omega^2  \cos\theta\,  H^2 \rho_{0}^2  \mu_H'' \right) = 0,
\end{multline}
\end{subequations}
  where primes now denote $z$-derivatives. A non-trivial numerical test of the equations is provided by the observation that the vertical component of the total wave energy flux is constant in $z$, as it should be.

\bsp	
\label{lastpage}
\end{document}